\documentclass[twoside,twocolumn,9pt]{article}
\usepackage{extsizes}
\usepackage[super,sort&compress,comma]{natbib} 
\usepackage[version=3]{mhchem}
\usepackage[left=1.5cm, right=1.5cm, top=1.785cm, bottom=2.0cm]{geometry}
\usepackage{balance}
\usepackage{widetext}
\usepackage{times,mathptmx}
\usepackage{sectsty}
\usepackage{graphicx} 
\usepackage{lastpage}
\usepackage[format=plain,justification=raggedright,singlelinecheck=false,font={stretch=1.125,small,sf},labelfont=bf,labelsep=space]{caption}
\usepackage{float}
\usepackage{fancyhdr}
\usepackage{fnpos}
\usepackage[english]{babel}
\usepackage{array}
\usepackage{droidsans}
\usepackage{charter}
\usepackage[T1]{fontenc}
\usepackage[usenames,dvipsnames]{xcolor}
\usepackage{setspace}
\usepackage[compact]{titlesec}

\usepackage{epstopdf}

\definecolor{cream}{RGB}{222,217,201}

\begin{document}

\pagestyle{fancy}
\thispagestyle{plain}
\fancypagestyle{plain}{

\fancyhead[C]{\includegraphics[width=18.5cm]{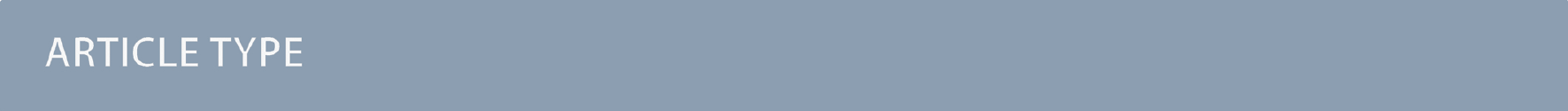}}
\fancyhead[L]{\hspace{0cm}\vspace{1.5cm}\includegraphics[height=30pt]{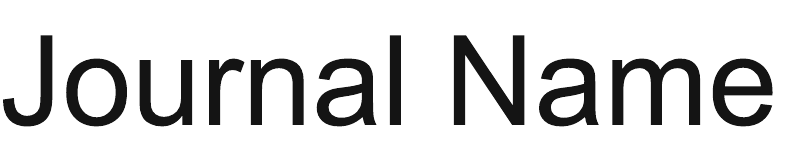}}
\fancyhead[R]{\hspace{0cm}\vspace{1.7cm}\includegraphics[height=55pt]{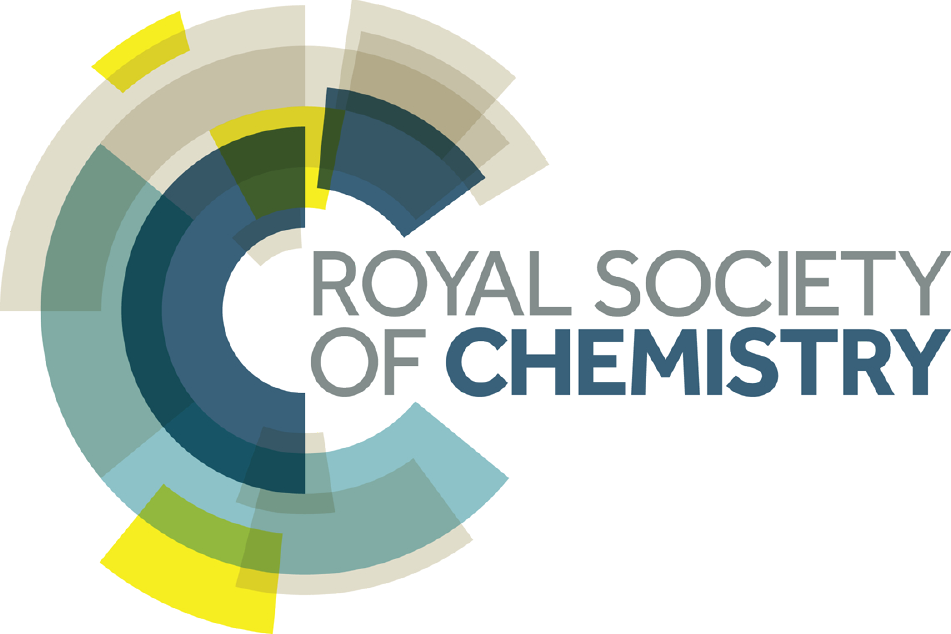}}
\renewcommand{\headrulewidth}{0pt}
}

\makeFNbottom
\makeatletter
\renewcommand\LARGE{\@setfontsize\LARGE{15pt}{17}}
\renewcommand\Large{\@setfontsize\Large{12pt}{14}}
\renewcommand\large{\@setfontsize\large{10pt}{12}}
\renewcommand\footnotesize{\@setfontsize\footnotesize{7pt}{10}}
\makeatother

\renewcommand{\thefootnote}{\fnsymbol{footnote}}
\renewcommand\footnoterule{\vspace*{1pt}%
\color{cream}\hrule width 3.5in height 0.4pt \color{black}\vspace*{5pt}} 
\setcounter{secnumdepth}{5}

\makeatletter 
\renewcommand\@biblabel[1]{#1}            
\renewcommand\@makefntext[1]%
{\noindent\makebox[0pt][r]{\@thefnmark\,}#1}
\makeatother 
\renewcommand{\figurename}{\small{Fig.}~}
\sectionfont{\sffamily\Large}
\subsectionfont{\normalsize}
\subsubsectionfont{\bf}
\setstretch{1.125} 
\setlength{\skip\footins}{0.8cm}
\setlength{\footnotesep}{0.25cm}
\setlength{\jot}{10pt}
\titlespacing*{\section}{0pt}{4pt}{4pt}
\titlespacing*{\subsection}{0pt}{15pt}{1pt}

\fancyfoot{}
\fancyfoot[LO,RE]{\vspace{-7.1pt}\includegraphics[height=9pt]{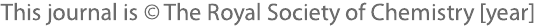}}
\fancyfoot[CO]{\vspace{-7.1pt}\hspace{13.2cm}\includegraphics{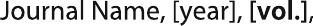}}
\fancyfoot[CE]{\vspace{-7.2pt}\hspace{-14.2cm}\includegraphics{head_foot/RF}}
\fancyfoot[RO]{\footnotesize{\sffamily{1--\pageref{LastPage} ~\textbar  \hspace{2pt}\thepage}}}
\fancyfoot[LE]{\footnotesize{\sffamily{\thepage~\textbar\hspace{3.45cm} 1--\pageref{LastPage}}}}
\fancyhead{}
\renewcommand{\headrulewidth}{0pt} 
\renewcommand{\footrulewidth}{0pt}
\setlength{\arrayrulewidth}{1pt}
\setlength{\columnsep}{6.5mm}
\setlength\bibsep{1pt}

\makeatletter 
\newlength{\figrulesep} 
\setlength{\figrulesep}{0.5\textfloatsep} 

\newcommand{\topfigrule}{\vspace*{-1pt}%
\noindent{\color{cream}\rule[-\figrulesep]{\columnwidth}{1.5pt}} }

\newcommand{\botfigrule}{\vspace*{-2pt}%
\noindent{\color{cream}\rule[\figrulesep]{\columnwidth}{1.5pt}} }

\newcommand{\dblfigrule}{\vspace*{-1pt}%
\noindent{\color{cream}\rule[-\figrulesep]{\textwidth}{1.5pt}} }

\makeatother

\twocolumn[
  \begin{@twocolumnfalse}
\vspace{3cm}
\sffamily
\begin{tabular}{m{4.5cm} p{13.5cm} }

\includegraphics{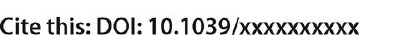} & \noindent\LARGE{\textbf{Shapes for maximal coverage for two-dimensional random sequential adsorption}} \\
\vspace{0.3cm} & \vspace{0.3cm} \\

 & \noindent\large{Micha\l{} Cie\'sla,$^{\ast}$\textit{$^{a}$} Grzegorz Paj\k{a}k,\textit{$^{a\dag}$} and Robert M. Ziff\textit{$^{b\ddag}$}} \\

\includegraphics{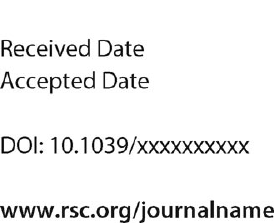} & \noindent\normalsize{The random sequential adsorption of various particle shapes is studied in order to determine the influence of particle anisotropy on the saturated random packing. For all tested particles there is an optimal level of  anisotropy which maximizes the saturated packing fraction. It is found that a concave shape derived from a dimer of disks gives a packing fraction of $0.5833$, which is comparable to the maximum packing fraction of ellipsoids and spherocylinders and higher than other studied shapes.  Discussion why this shape is beneficial for random sequential adsorption is given.} \\

\end{tabular}

 \end{@twocolumnfalse} \vspace{0.6cm}

  ]

\renewcommand*\rmdefault{bch}\normalfont\upshape
\rmfamily
\section*{}
\vspace{-1cm}


\footnotetext{\textit{$^{a}$~M.\ Smoluchowski Institute of Physics, Department of Statistical Physics, Jagiellonian University, \L{}ojasiewicza 11, 30-348 Krak\'ow, Poland.}}
\footnotetext{\textit{$^{b}$~Center for the Study of Complex Systems and Department of Chemical Engineering, University of Michigan, Ann Arbor MI 48109-2136 USA.}}


\footnotetext{\textit{$\ast$~michal.ciesla@uj.edu.pl}}

\footnotetext{\textit{$\dag$~grzegorz@th.if.uj.edu.pl}}

\footnotetext{\textit{$\ddag$~rziff@umich.edu}}


%


%


%
%
%
\section{Introduction}
Random sequential adsorption (RSA) is a fundamental problem that has attracted unabated interest for decades.  In 1939, Flory \cite{bib:Flory1939} studied the attachment of blocking pendant groups on a linear polymer, which is effectively a one-dimensional RSA problem.  R\'enyi \cite{bib:Renyi1958} introduced another famous one-dimensional RSA problem---the parking of cars along an unmarked curb.  
Feder \cite{bib:Feder1980} helped to make RSA a very popular tool for modeling monolayers obtained as a result of irreversible adsorption \cite{bib:Adamczyk2012, bib:Dabrowski2001,bib:Evans1993}.  More recently, random packings generated by RSA have been of interest in a number of scientific fields, e.g., soft matter \cite{bib:Torquato2010, bib:Corwin2010, bib:Kyrylyuk2011},
surface science \cite{bib:EvansLiu2014}, mathematics \cite{bib:Zong2014}, telecommunication \cite{bib:Hastings2005} and information theory \cite{bib:Coffman1998}.

The RSA algorithm is based on consecutive tries to add a particle to a packing. Firstly, a  particle's position (and orientation in case of anisotropic objects) is drawn according to the probability distribution which reflects the structure of an underlying surface -- a homogeneous surface corresponds to the uniform probability distribution. Then the particle is tested if it overlaps or intersects with any of particles already added to the packing. If not, the particle is added to the packing; otherwise it is abandoned. These steps should be continued until the packing is saturated, i.e., there is no room for any additional particle on a surface. However, typically an algorithm is stopped when the probability of successful adding of a particle is sufficiently small, and extrapolations are made to estimate the maximum ``jamming" coverage.

The properties of packings generated by RSA algorithm have been checked for a number of different particle shapes, e.g.,\ spheres \cite{bib:Feder1980,bib:Ciesla2013ms,bib:Zhang2013}, spherocylinders and ellipsoids \cite{bib:Sherwood1990,bib:Viot1992}, rectangles \cite{bib:Vigil1989,bib:Vigil1990}, and polymers \cite{bib:Ciesla2013pol,bib:Shelke2015}. Results obtained for anisotropic particles show that saturated random coverage fraction reaches its maximum for moderate anisotropy, i.e., when long-to-short particle axis ratio is approximately $1.5$-$2.0$ \cite{bib:Viot1992,bib:Vigil1990}.  The highest saturated coverage fraction obtained was $\theta_\mathrm{max} = 0.583 \pm 0.001$, for both ellipsoids and spherocylinders \cite{bib:Sherwood1990,bib:Viot1992}. Both of these are convex shapes. The highest saturated coverage fraction for concave particles was observed for a dimer built of two overlapping disks, with $\theta_\mathrm{max} = 0.5793 \pm 0.0001$ \cite{bib:Ciesla2014dim}, slightly lower than above values.

A similar problem of finding the highest possible random close packing of various objects in three dimensions is crucial for understanding granular and soft matter properties.  \cite{bib:Baule2014}. However close packings differs significantly from saturated packings generated by the RSA algorithm as in the first case particles are tightly packed and are touching most of their closest neighbors.  It is interesting that the random close packing fraction of ellipsoids can be as much as $0.74$ while for spheres it is $0.64$ \cite{bib:Donev2004,bib:Man2005}. Moreover, it occurs that the highest packing were found for particles (ellipsoids, spherocylinders or dimers) of long-to-short axis ratio around $1.5$, which is similar as for RSA in two dimensions. These results were confirmed both: numerically \cite{bib:Donev2004,bib:Faure2009,bib:Zhao2012} and theoretically \cite{bib:Baule2013}.

The primary aim of this study is to analyze set of different concave shapes in order to find if one of them can give a higher saturated random packing fraction than previously found.  This will help in designing particles, such as nanoparticles, that can cover a surface most efficiently.   Additionally, we study the kinetics of RSA for nearly symmetric particles as they appear to be very sensitive on particle shape anisotropy \cite{bib:Ciesla2014dim, bib:Ciesla2014ring}.
\section{Model}
\begin{figure}[htb]
\centerline{%
\includegraphics[width=0.9\columnwidth]{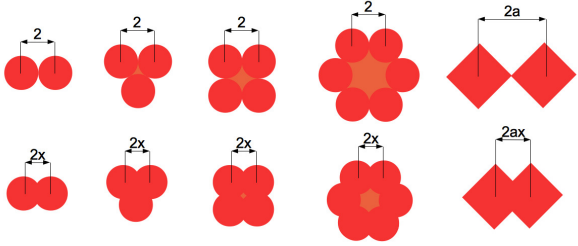}}
\caption{The five types of shapes for which saturated random packings are studied. Disks have a unit radius, and parameter $x \in[0,1]$ corresponds to half the distance between closest spheres or squares. Squares have a unit side size, thus parameter $a$ equals $\sqrt{2}/2$.}
\label{fig:shapes}
\end{figure}

The shapes we consider are shown in Fig.~\ref{fig:shapes}. 
Dimers, trimers, tetramers and hexamers are built of identical disks of unit radius, and the distance between neighboring disks centers is $2x$. The parameter $x$ if not stated otherwise was taken from the interval $[0,1]$.  For $x=0$ all these shapes are equal to a single disk, which of course is isotropic, while for $x=1$ each disk touches two neighboring disks as presented in the upper row of Fig.~\ref{fig:shapes}. Note that the space inside ring of disks also belongs to the particle. Such choice of shapes originates from \cite{bib:Ciesla2014dim}, where properties of dimer random packings were carefully studied. Here we want to test how the more complex shape of particle shape affects properties of saturated random packing. The last shape is built of two partially overlapping squares of a unit side size. Parameter $a$ shown in Fig.~\ref{fig:shapes} equals to $\sqrt{2}/2$. This shape is an analogue of generalized dimer but instead of disks, it is built of two squares.
The surface area, for $0 \leq x \leq 1$, of dimer ($S_2$), trimer ($S_3$), tetramer ($S_4$), hexamer ($S_6$) and the square dimer ($S_{2S}$) presented in Fig.~\ref{fig:shapes} are given respectively:
\begin{equation}
S_{2}(x) = 2 \pi - 2 \left( \arccos x - x \sqrt{1-x^2} \right) 
\end{equation}
\begin{eqnarray}
S_{3}(x) &=&  \pi(3 \mp 1)/2 + \arcsin x + \sqrt{3} x^2+4x \sqrt{1-x^2} \cr
 &+& \arcsin \left( \frac{1}{2} \left(x-\sqrt{3-3 x^2}\right) \right) \cr
 &\pm& \arcsin \left( \frac{1}{2} \left(x+\sqrt{3-3 x^2}\right) \right) \cr
 &\pm&  \frac{1}{4} x \sqrt{2 x^2-2 x\sqrt{3-3 x^2} +1} \cr
 &+&  \frac{1}{4} x \sqrt{2 x^2+2 x\sqrt{3-3 x^2} +1} \cr
 &\pm& \frac{\sqrt{3}}{4} \sqrt{\left(x^2-1\right) \left(-2 x^2+2x \sqrt{3-3 x^2}
   -1\right)} \cr
   &-&  \frac{\sqrt{3}}{4} \sqrt{\left(x^2-1\right) \left(-2 x^2-2x \sqrt{3-3 x^2}
   -1\right)}
\end{eqnarray}
%
%
\begin{equation}
S_4(x) = \pi + 4 x^2 + 4 x \sqrt{1 - x^2} + 4 \arcsin x
\end{equation}
\begin{eqnarray}
S_{6}(x) &=& 
\pi(3 \mp 3)/2 + 2 \arcsin x + 6 \sqrt{3} x^2+8x \sqrt{1-x^2} \cr
 &+& (1 \pm 1) \left[ \arcsin \left( \sqrt{3} x \right) +  \arcsin \left( \sqrt{1-3 x^2} \right) \right] \cr
 &+& 2 \arcsin \left( \frac{1}{2} \left(x-\sqrt{3-3 x^2}\right) \right) \cr
 &\pm& 2 \arcsin \left( \frac{1}{2} \left(x+\sqrt{3-3 x^2}\right) \right) \cr
 &\pm& \frac{1}{2} x \sqrt{2 x^2-2 x\sqrt{3-3 x^2} +1} \cr
 &+&\frac{1}{2} x \sqrt{2 x^2+2 x\sqrt{3-3 x^2} +1} \cr
 &\pm& \frac{\sqrt{3}}{2} \sqrt{\left(x^2-1\right) \left(-2 x^2+2x \sqrt{3-3 x^2} - 1\right)} \cr 
   &-& \frac{\sqrt{3}}{2} \sqrt{\left(x^2-1\right) \left(-2 x^2-2x \sqrt{3-3 x^2} -1 \right) }
\end{eqnarray}
\begin{equation}
S_{2S}(x) = 2\left(1+2x-x^2\right), 
\end{equation}
where in $S_3(x)$ and $S_6(x)$, the upper sign in $\pm$ or $\mp$ is for $0 \le x < 1/2$, and the lower sign is for $1/2  \le x \le 1$. 
\par
These shapes were thrown onto a square surface of a side size up to $1000$ with an area up to $S_C=10^6$. We decided to use open boundary conditions similar to \cite{bib:Ciesla2014dim}. Such a choice allowed us to use data published there for comparison purposes. The packing fraction is equal to:
\begin{equation}
\label{eq:nq}
\theta(t) = N(t) \frac{S_p}{S_C} = \rho(t) S_p,
\end{equation}
where $S_p$ is a particle surface area and $N(t)$ is a number of particles in a packing after a number of RSA iterations corresponding to time $t$ measured in the dimensionless time units 
\begin{equation}
t = n \frac{S_p}{S_C},
\end{equation}
where $n$ is number of RSA algorithm steps, and $\rho(t) = N(t)/S_C$ is the surface density of particles in a packing.
\par
The simulation was stopped when $t=10^5$. At this moment, the probability that a randomly chosen place is large enough for an additional particle is well below $10^{-6}$. The number of particles in a single random packing was at the order of $10^5$. To improve statistics, up to $100$ independent simulations were performed for each shape. During the simulations the number of particles in a packing $N(t)$ was recorded. 
\section{Results}
\begin{figure}[htb]
\centerline{%
\includegraphics[width=0.45\columnwidth]{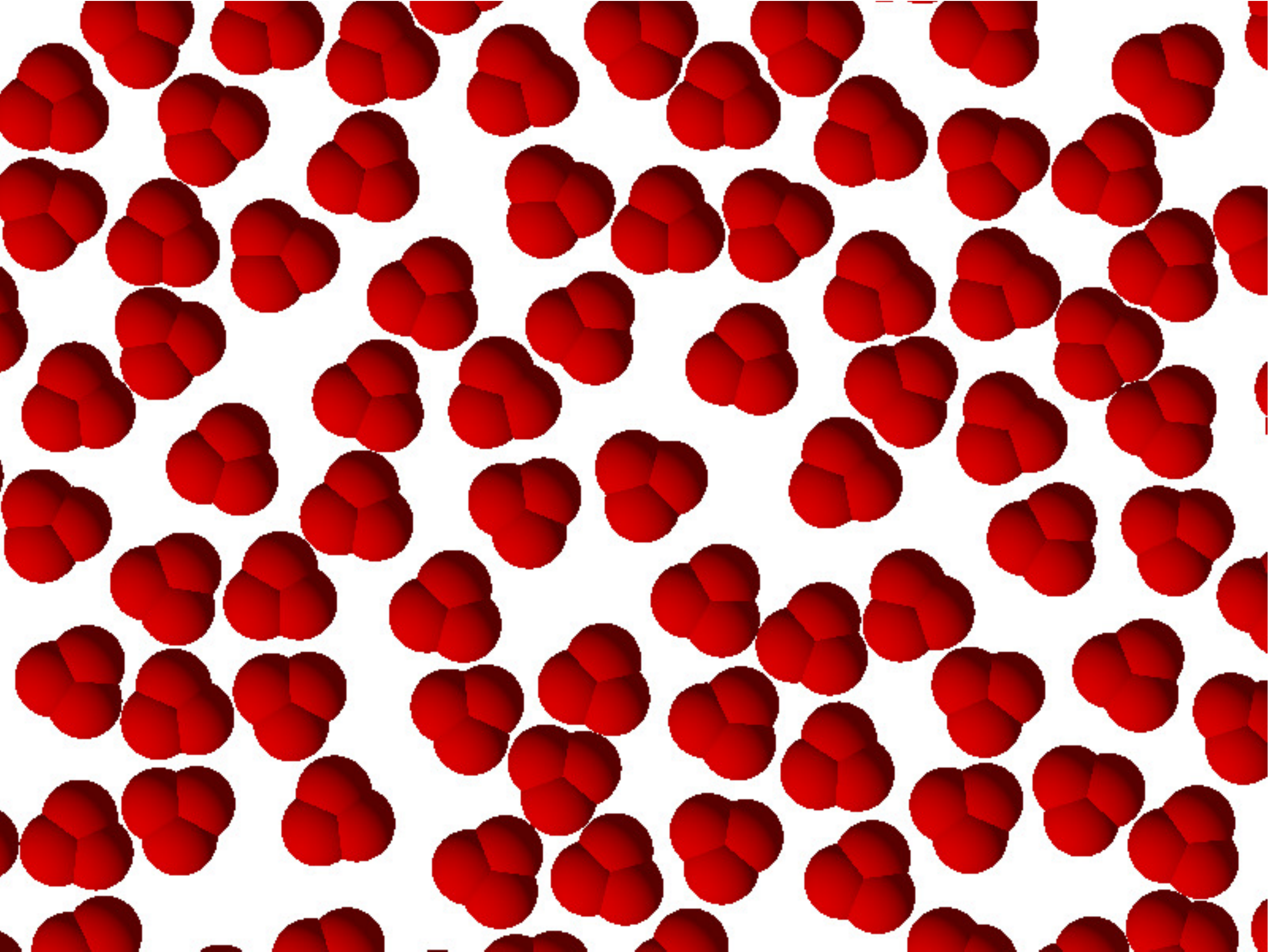}
\hspace{0.05\columnwidth}
\includegraphics[width=0.45\columnwidth]{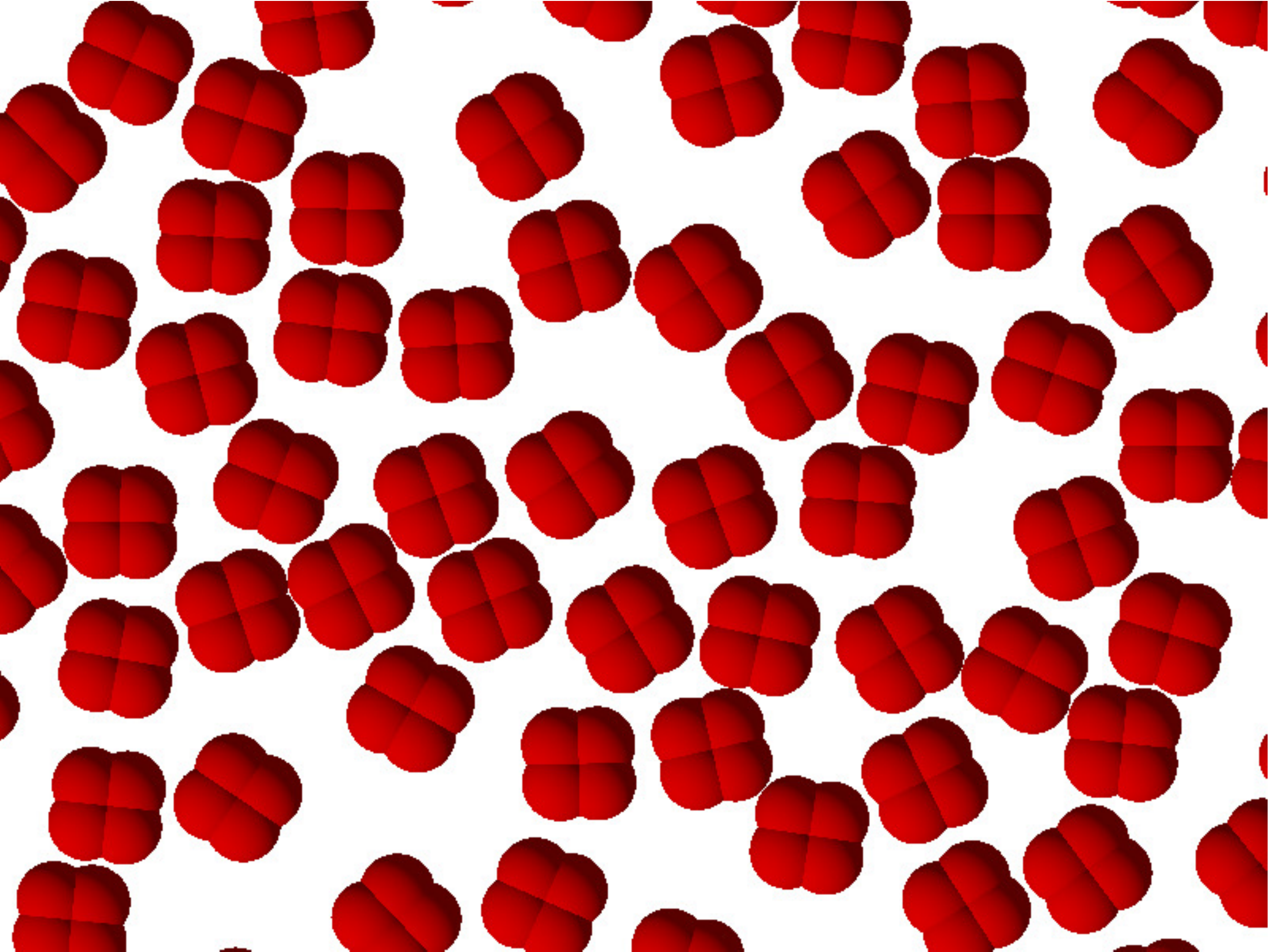}
}
\vspace{0.05\columnwidth}
\centerline{%
\includegraphics[width=0.45\columnwidth]{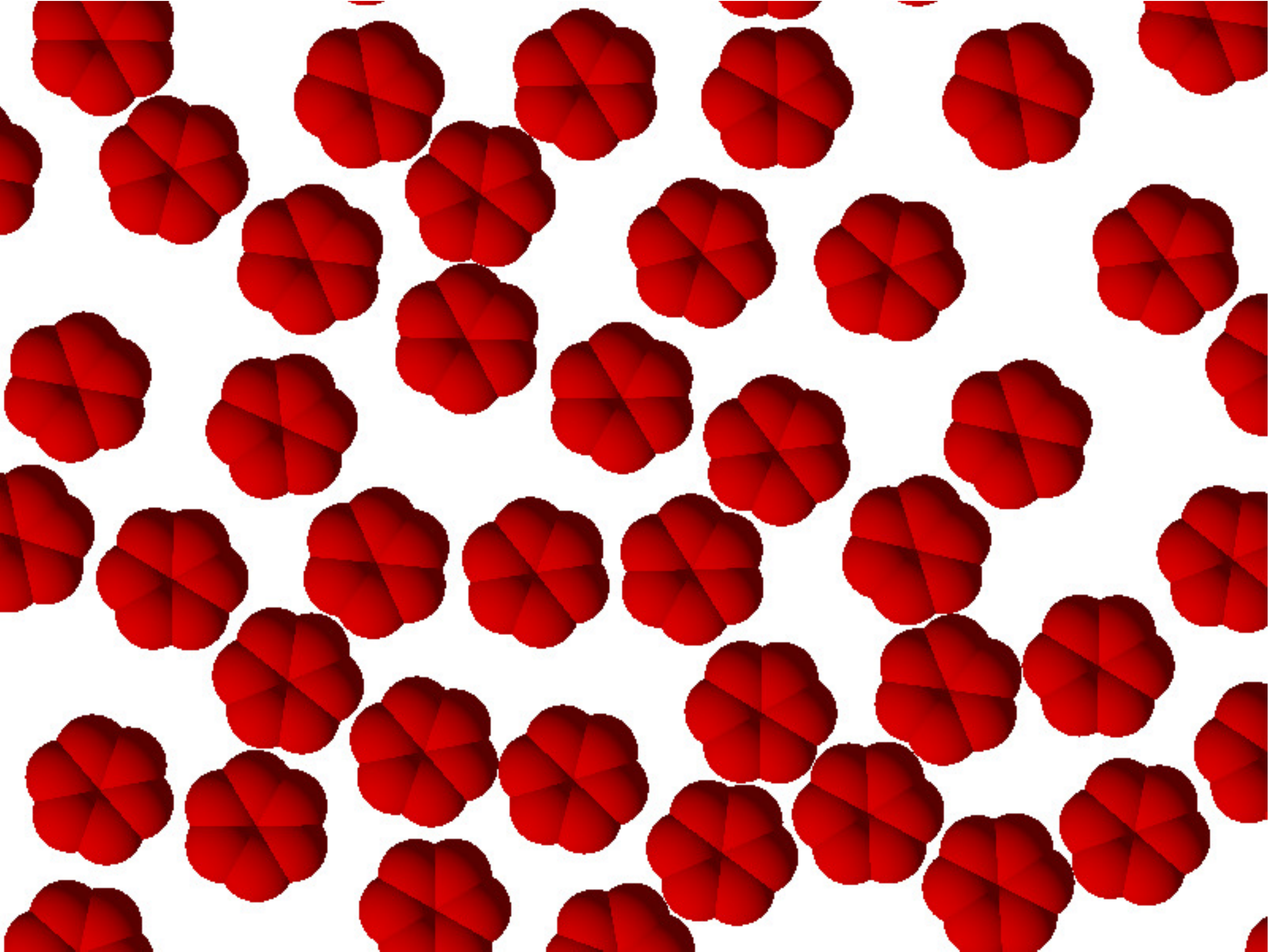}
\hspace{0.05\columnwidth}
\includegraphics[width=0.45\columnwidth]{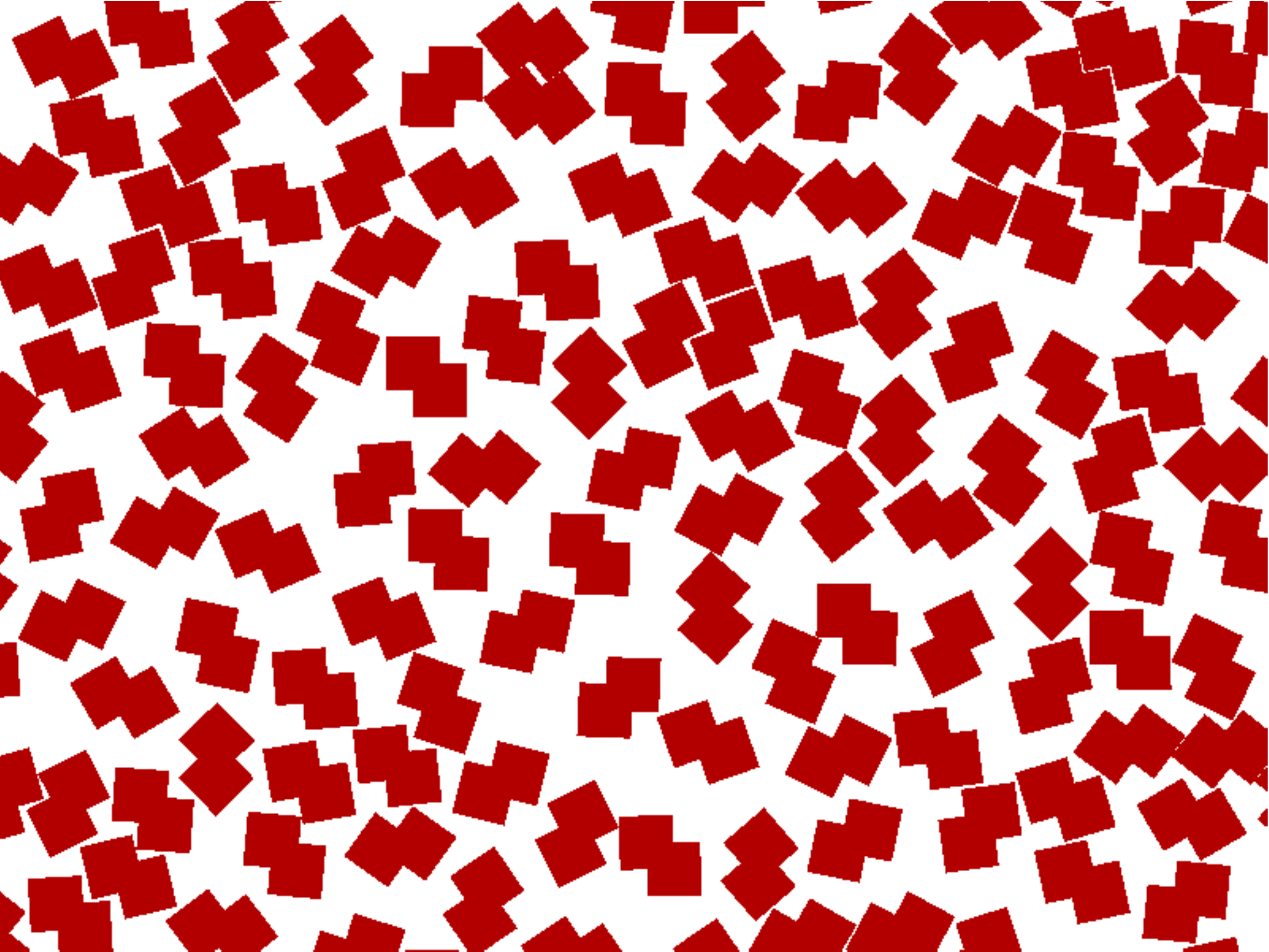}
}\caption{Examples of obtained random packings for studied shapes after $t=10^5$ iterations of RSA algorithm. Parameter $x$ equals $0.5$ for all four shapes. At this point the systems were at approximately 99\% of the extrapolated maximum packing.}
\label{fig:examples}
\end{figure}

Fragments of sample packings are presented in Fig.~\ref{fig:examples}.
Examples of dimer packings have been shown in \cite{bib:Ciesla2014dim}. The total number of RSA algorithm steps is large, but finite. Thus, the resulting packings may be not be saturated. To find number of particles in saturated packing from a finite-time simulation a knowledge about RSA kinetics is essential. Therefore, to study saturated packing fractions the kinetics of the process has to be clarified first.
\subsection{RSA kinetics}
The RSA kinetics for large enough time $t$ is governed by the power law \cite{bib:Pomeau1980, bib:Swendsen1981}
\begin{equation}
\label{eq:fl}
 \theta(t) = \theta_\mathrm{max} - A t^{-1/d}
\end{equation}
where $\theta_\mathrm{max} \equiv \theta(t\to\infty)$, $A$ is a positive constant and $d$ depends on particle shape and properties of a surface. For flat and homogeneous surfaces $d$ can be interpreted as a number of degrees of freedom of a particle \cite{bib:Ciesla2013pol, bib:Hinrichsen1986}. Thus, for RSA of disks on two dimensional flat surface $d=2$ but for anisotropic particles $d=3$ as an orientation of a particle is an additional degree of freedom. It is worth noting, that even when particle anisotropy is very small, $d=3$ describes the ultimate asymptotic behavior \cite{bib:Viot1992, bib:Ciesla2014ring, bib:Ciesla2014dim}. 
\par
\begin{figure}[htb]
\centerline{%
\includegraphics[width=0.45\columnwidth]{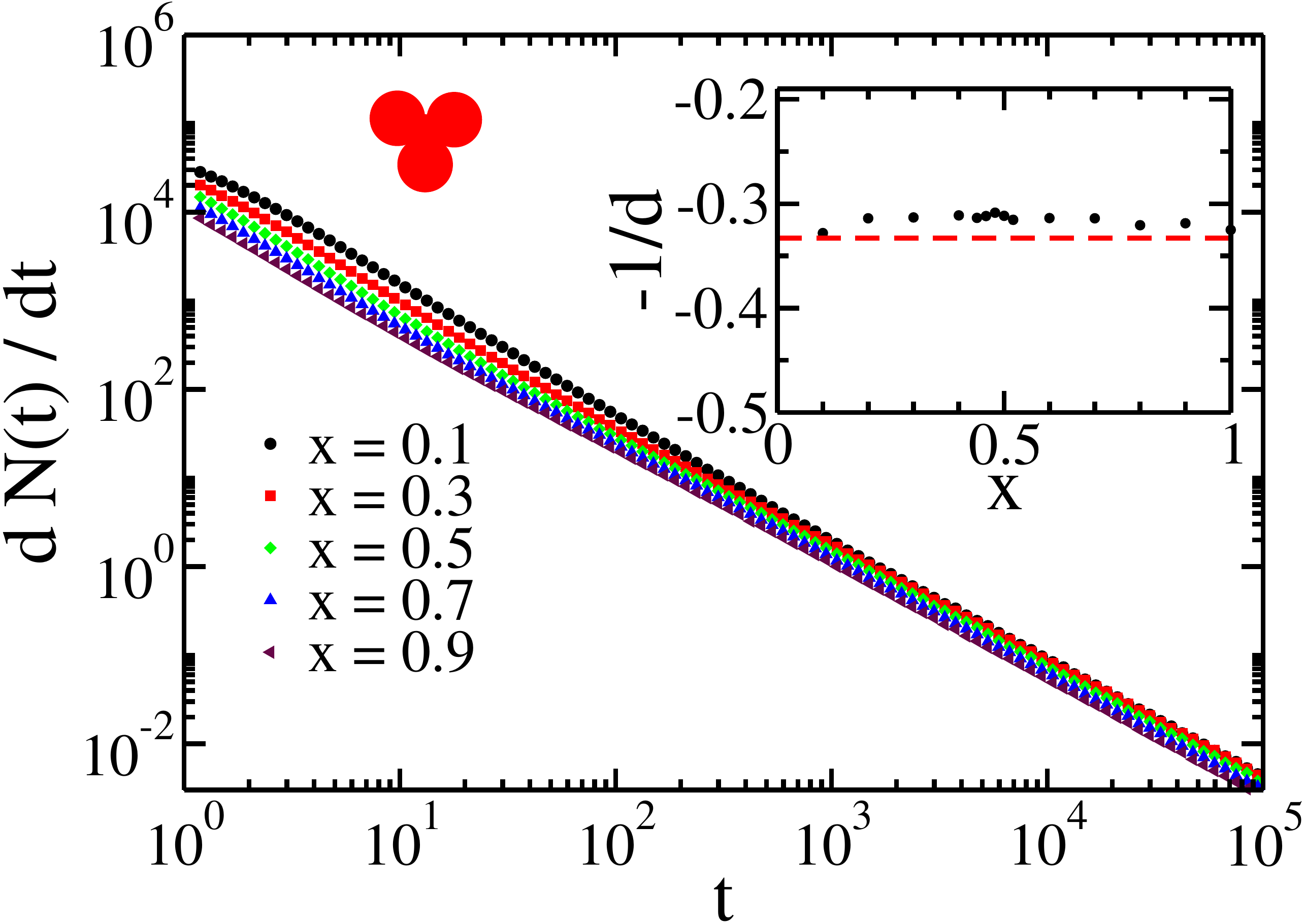}
\hspace{0.05\columnwidth}
\includegraphics[width=0.45\columnwidth]{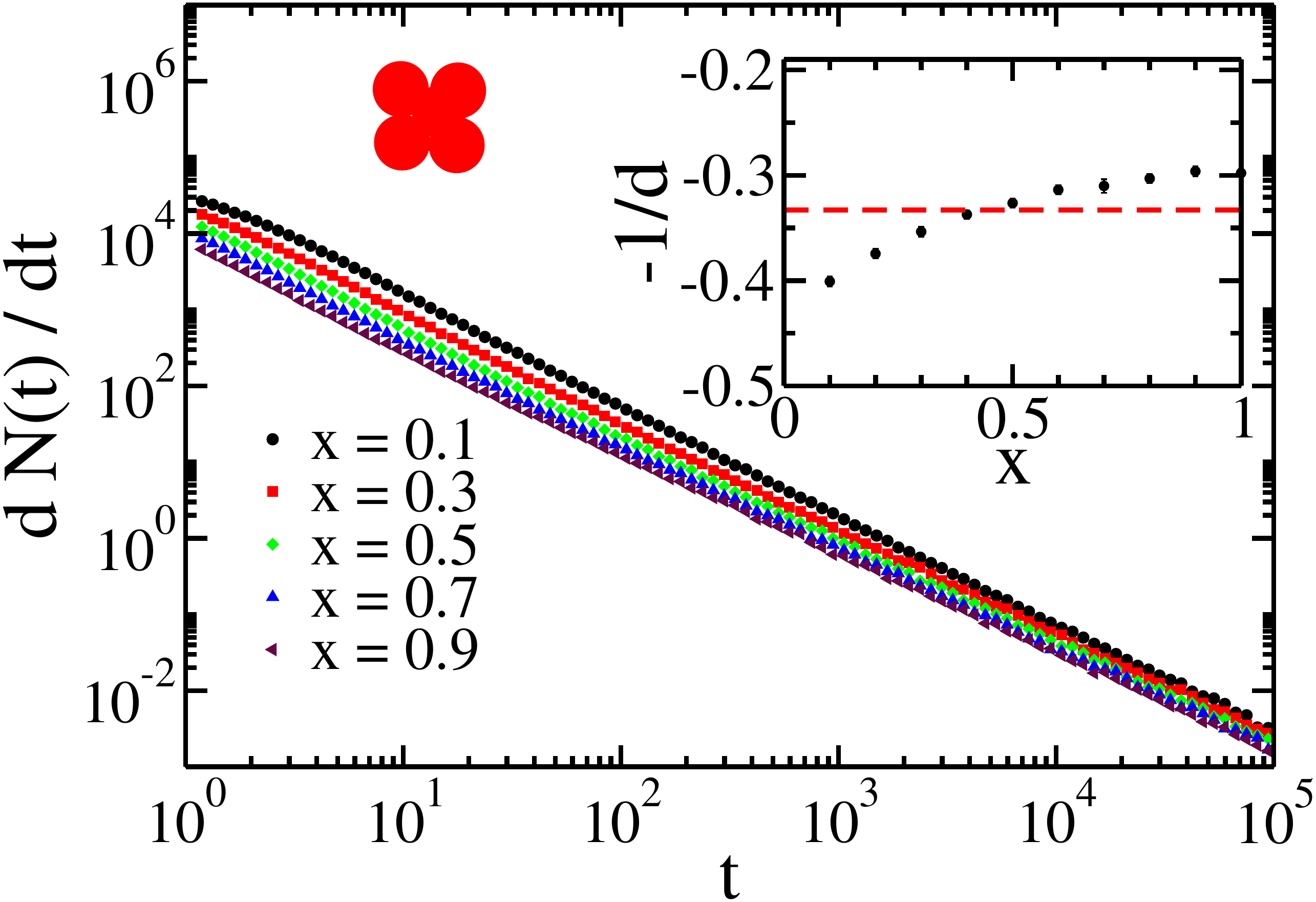}
}
\vspace{0.05\columnwidth}
\centerline{%
\includegraphics[width=0.45\columnwidth]{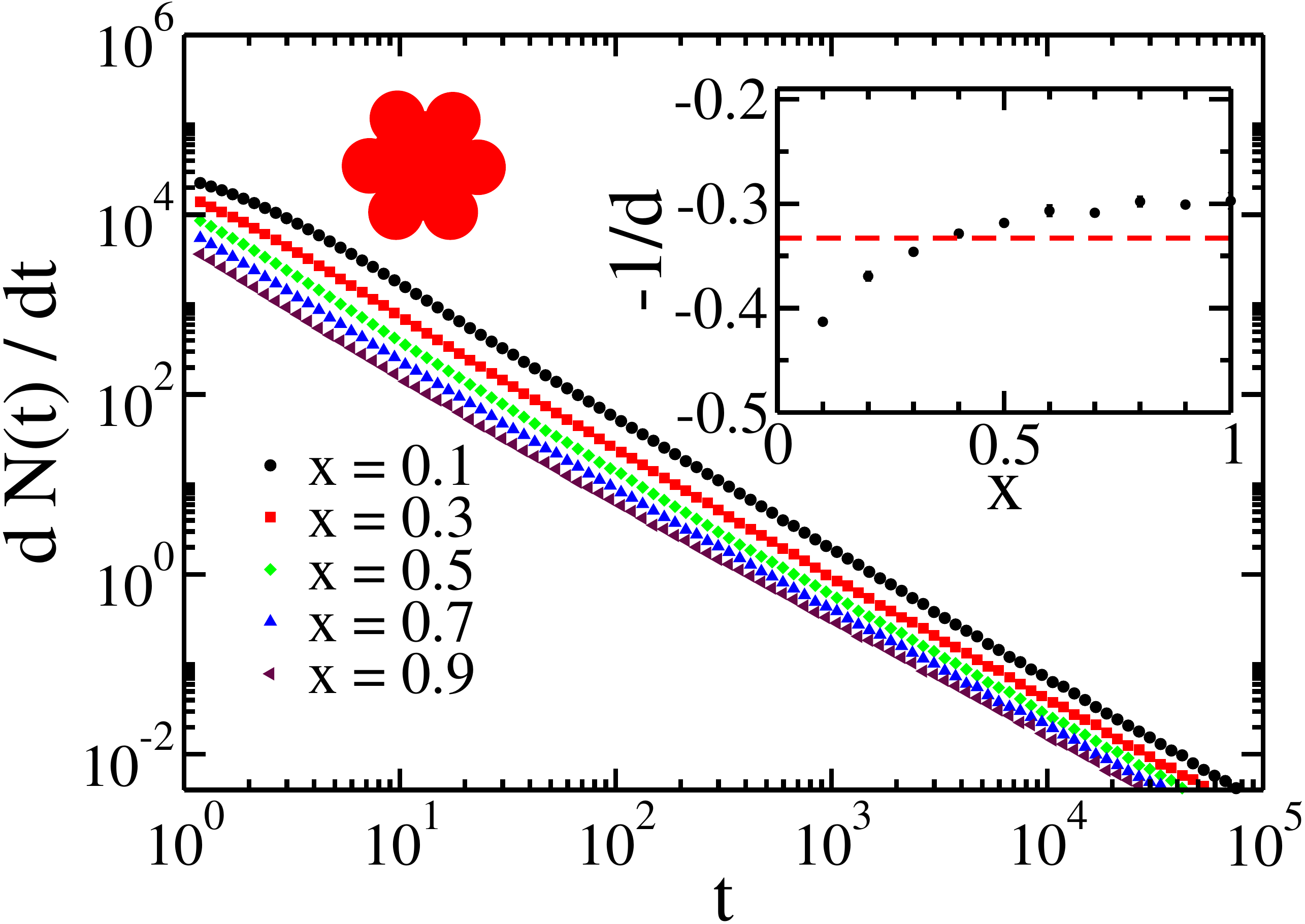}
\hspace{0.05\columnwidth}
\includegraphics[width=0.45\columnwidth]{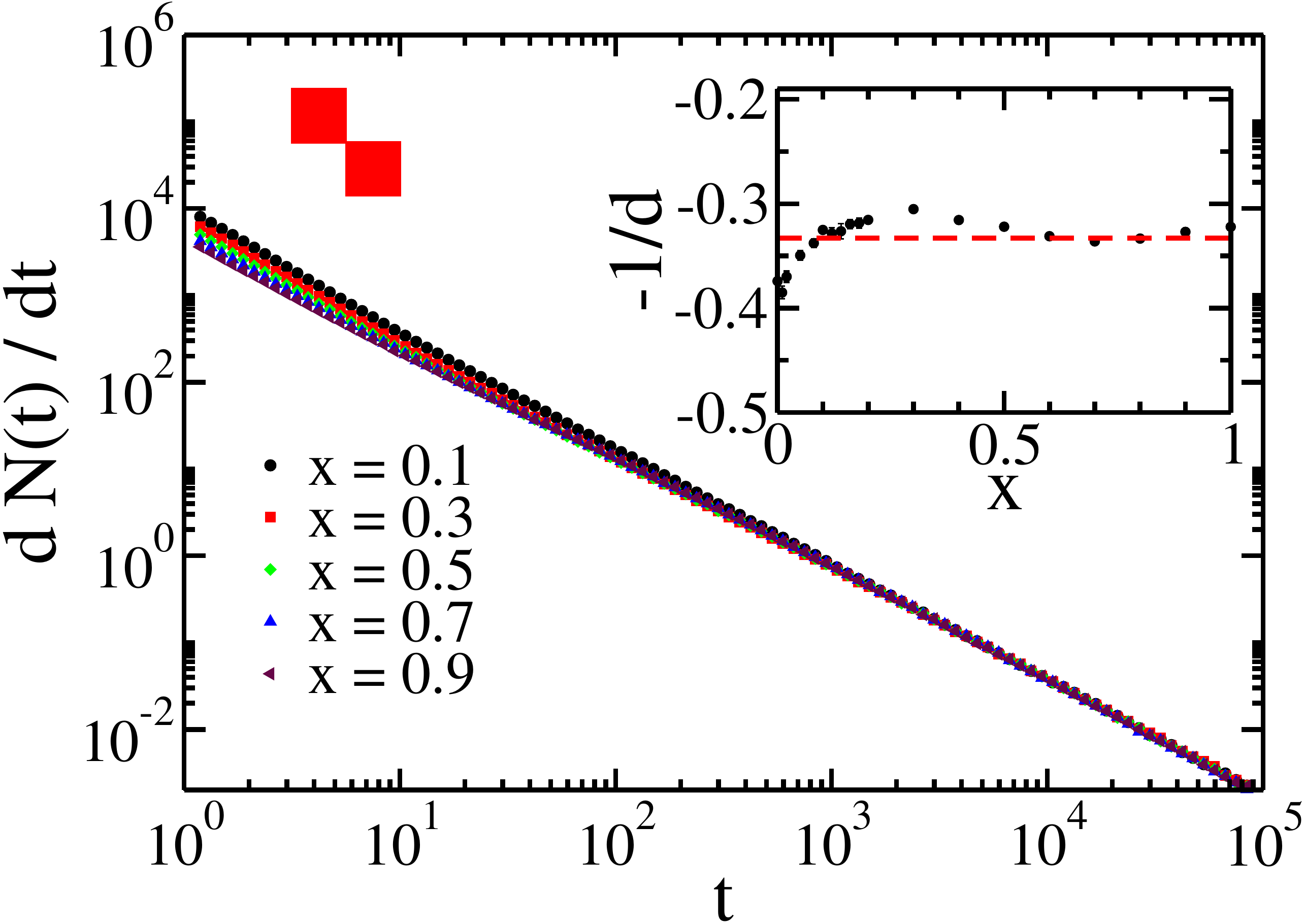}
}
\caption{(Color online) RSA kinetics for trimers, tetramers, hexamers and square dimers. Main panels show number of added particles in a time unit versus time. Insets show fitted exponent in Eq.~\ref{eq:fl} dependence on $x$. Error bars are smaller than symbol sizes. Dashed lines correspond to the $d=3$ characteristic behavior for anisotropic molecules.}
\label{fig:kinetics}
\end{figure}

The RSA kinetics of our shapes are presented in Fig.~\ref{fig:kinetics}.
Data for dimers has been presented in \cite{bib:Ciesla2014dim}. For all shapes Eq.~(\ref{eq:fl}) is fulfilled as the data obtained from numerical simulations lays along straight lines in a log-log scale. The parameter $d$ obtained from fitting numerical data to the relation (\ref{eq:fl}) is, as expected for anisotropic particles, around $3$. In a limit of small $x$, for tetramers and hexamers the drop of $d$ down to $2$ is observed, which is expected, as the particle shape arises from disk-like geometry and the simulations have not yet reached the true asymptotic behavior.  In other words, the apparent change of $d$ from 3 to 2 as $x$ decreases represents crossover behavior between the two exponents. For trimers this drop occurs for very small $x$, which resembles its behavior for dimers \cite{bib:Ciesla2014dim}. In the case of square dimers very slight decrease is observed, because even for $x=0$ the particle is still anisotropic. This is in agreement with previous observations \cite{bib:Viot1990breakdown, bib:Ciesla2014stars}.
\subsection{Saturated random packing fractions}
\label{sec:packing}
\begin{figure}[htb]
\centerline{%
\includegraphics[width=0.7\columnwidth]{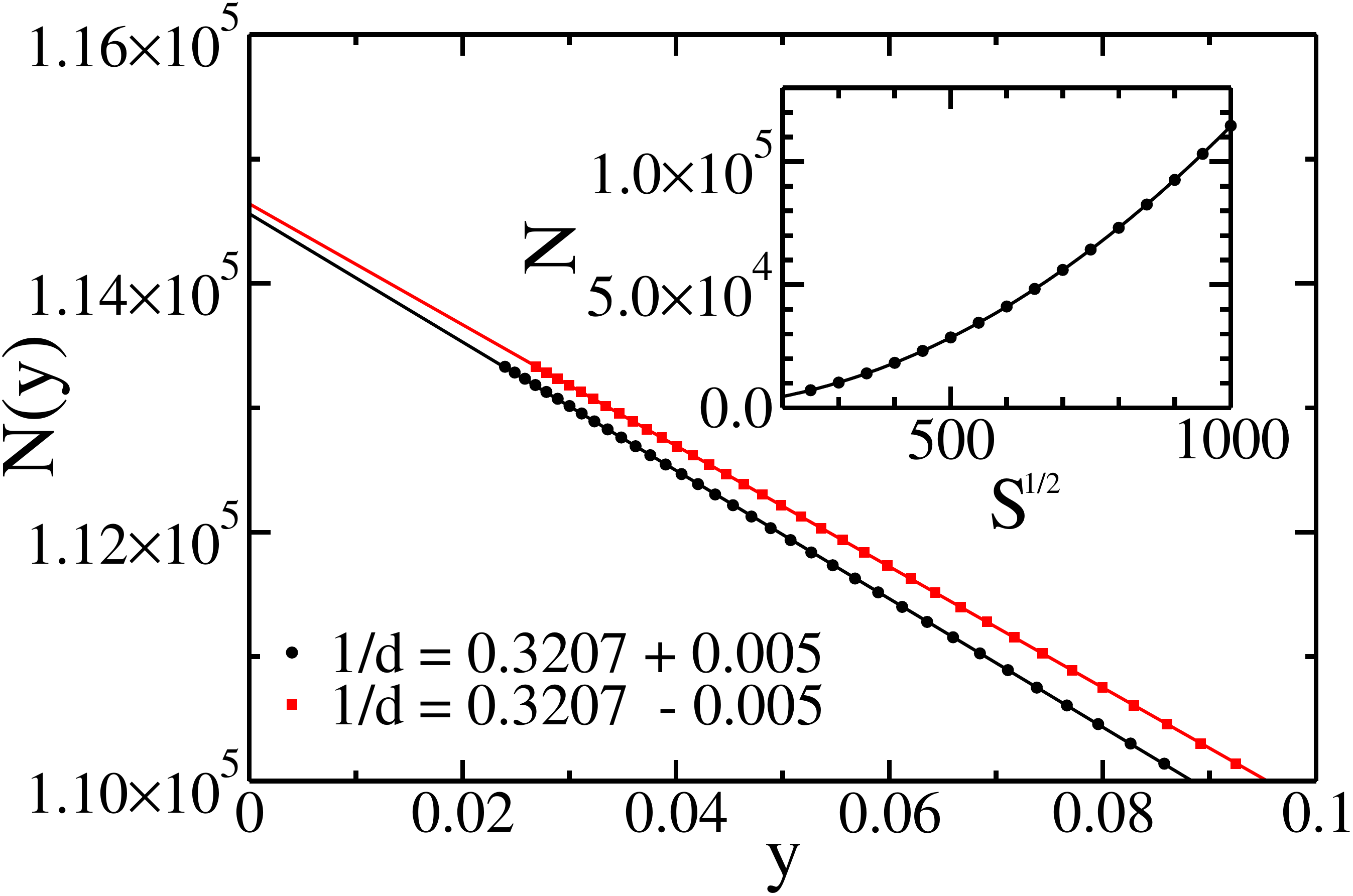}
}
\caption{(Color online) Estimation of number of particles in a saturated random packing. The main plot shows influence of $1/d$ exponent error on obtained number of particles in saturated packing $N_\mathrm{max}$ for dimer characterized by $x=0.5$. Dots represent data taken from simulations for $y=t^{-1/d \pm 0.005}$. Lines are linear fits: $N(y) = 1.1464\cdot 10^5 - 48598 y$ and $N(y) = 1.1456\cdot 10^5 - 51573 y$. Inset shows dependence of $N_\mathrm{max}$ on surface size. Dots represent data taken from simulations while line is a quadratic fit: $N = 162.64 - 0.70174 \sqrt{S} + 0.11499 S$.}
\label{fig:errors}
\end{figure}

Knowing parameter $d$ and substituting $y = t^{-1/d}$, Eq.~(\ref{eq:fl}) can be rewritten in a form: $N(y) = N_\mathrm{max} + A' y$. Thus, points $(N(y), y)$ measured during an RSA simulation lie along a straight line which crosses the axis $y=0$ at $N_\mathrm{max}$. It solves the problem of finite time simulations. The error of such $N_\mathrm{max}$ estimation originates in error of the exponent $-1/d$, which in our simulation is below $0.005$ (see Fig.\ref{fig:kinetics}). The corresponding component to the relative error of  $N_\mathrm{max}$ is $0.06\%$ (see Fig.\ref{fig:errors}).
Its value is comparable with the statistical error coming from averaging over $100$ different simulations. It is worth mentioning that the error component coming from finite time simulations can be totally eliminated for spherical particles using an algorithm presented in \cite{bib:Zhang2013}, which allows one to obtain saturated packings in a finite-time simulations by tracing regions where addition of particles is possible. Thus, the RSA sampling can be limited to these regions only. Unfortunately, in the case of anisotropic particles, the complexity of this algorithm raises as the success of placing a particle depends not only of its position but also on its orientation. Another problem originates in the finite size of a system, which is especially important as periodic boundary conditions are not used here. It is expected that in general the number of particles in a packing are a quadratic function of the surface size: $N(S_C) = a S_C + b\sqrt{S_C} + c$ (see the inset in Fig.\ref{fig:errors}). 
Thus for an infinite system ($S_C \to \infty$), a packing density $N(S_C)/S_C = a$. Therefore, to get a packing density in a limit infinitely large systems, several different sized systems should be analyzed to get the $a$ coefficient in the above relation. In this study $S_C$ varies from $2.5 \cdot 10^5$ up to $10^6$. Results presented in Fig.\ref{fig:errors} suggest that a systematic relative error coming from a finite surface size is in our case at the order of $0.1\%$ and is the biggest among  all error components. Therefore, the best way to improve the quality of obtained results is to simulate RSA on larger surfaces.

\begin{figure}[htb]
\centerline{%
\includegraphics[width=0.45\columnwidth]{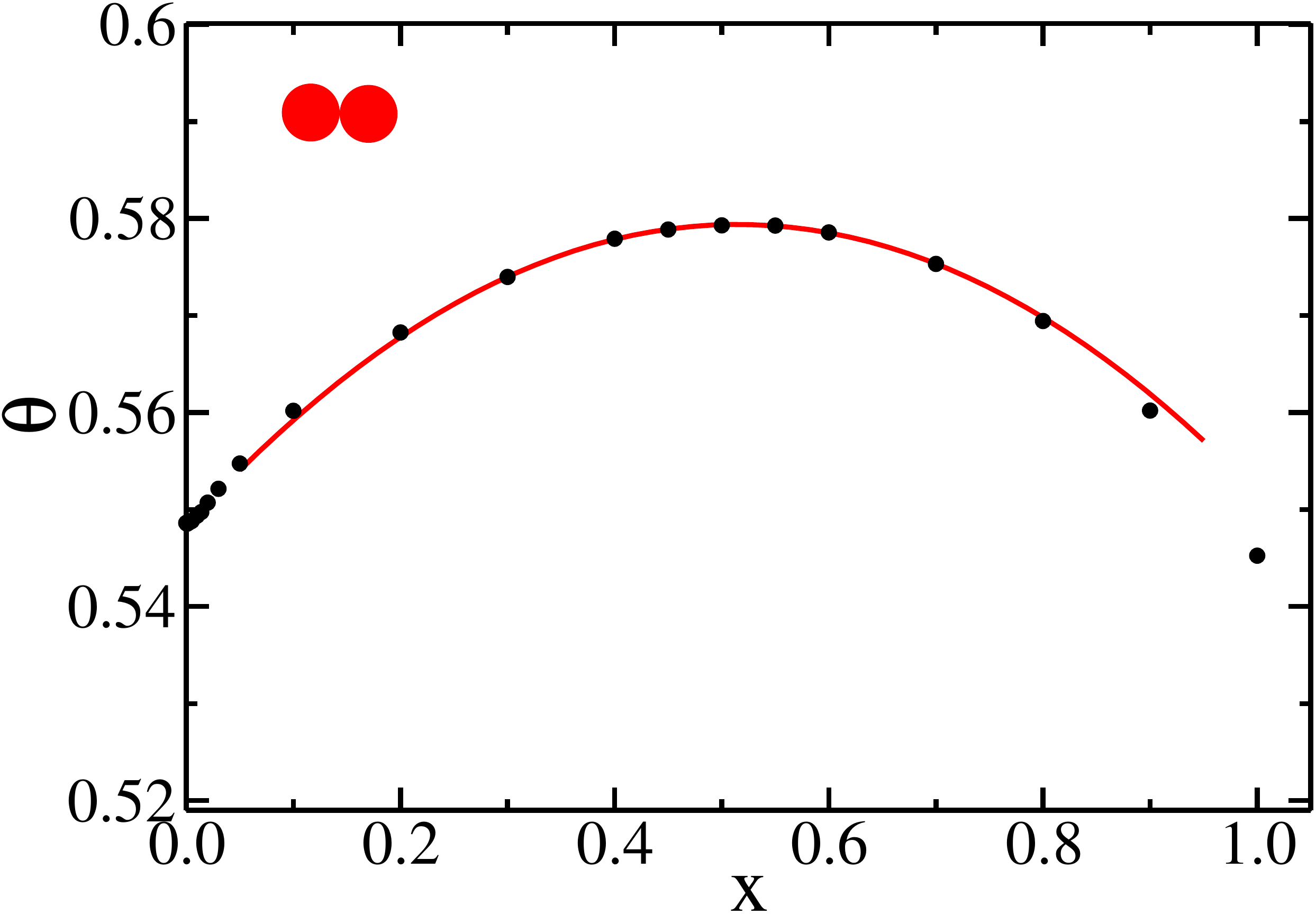}
\hspace{0.05\columnwidth}
\includegraphics[width=0.45\columnwidth]{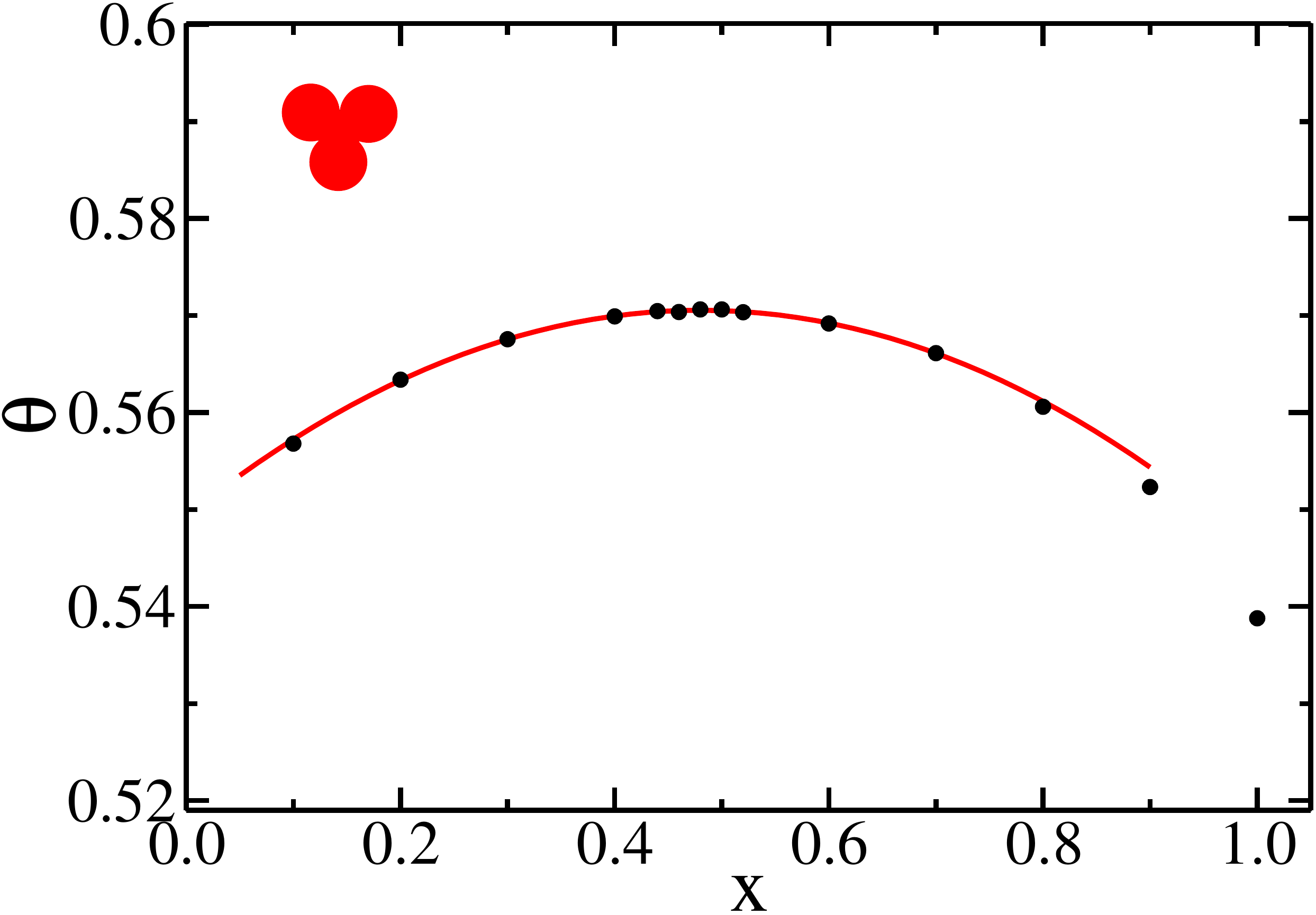}
}
\vspace{0.05\columnwidth}
\centerline{%
\includegraphics[width=0.45\columnwidth]{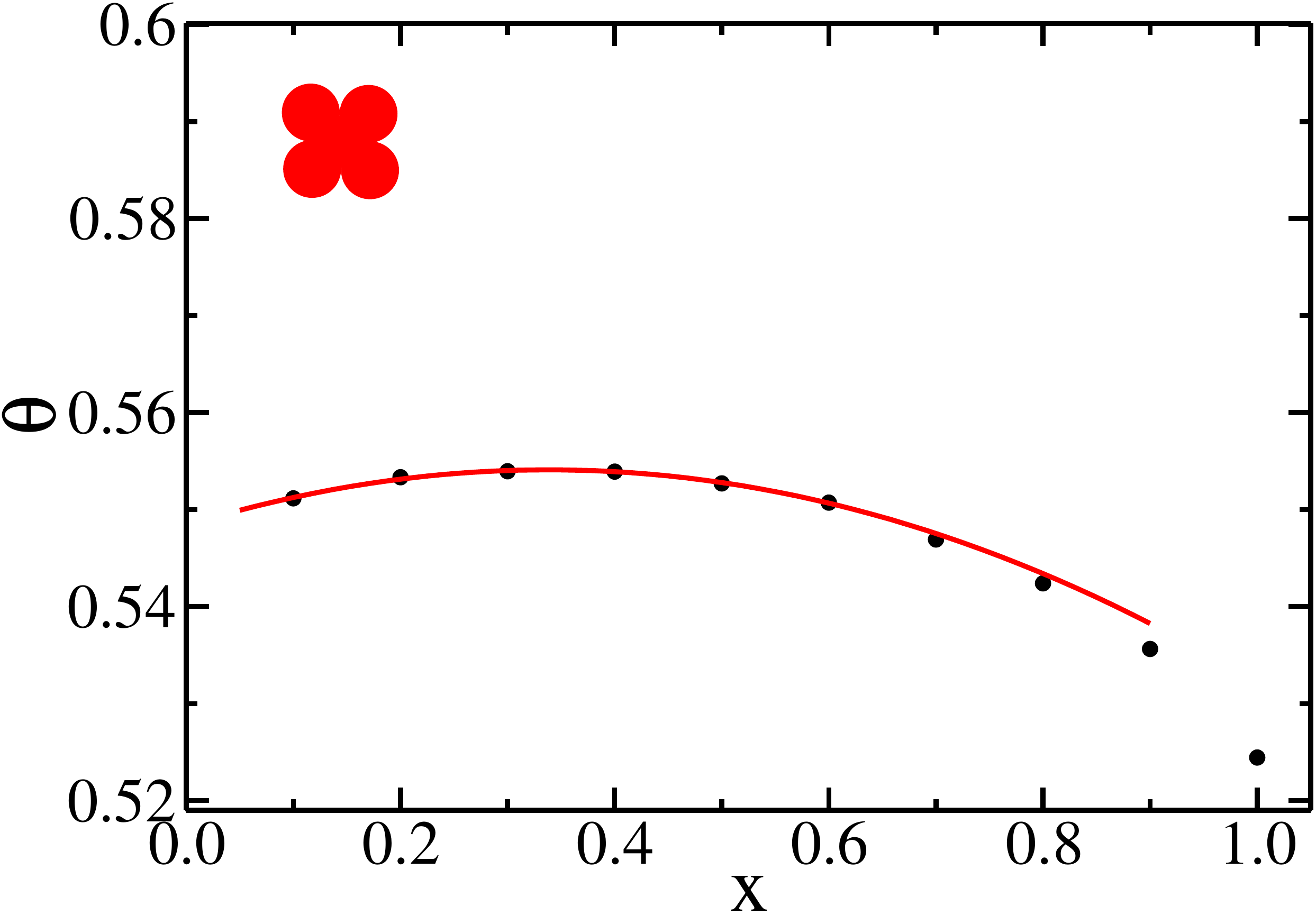}
\hspace{0.05\columnwidth}
\includegraphics[width=0.45\columnwidth]{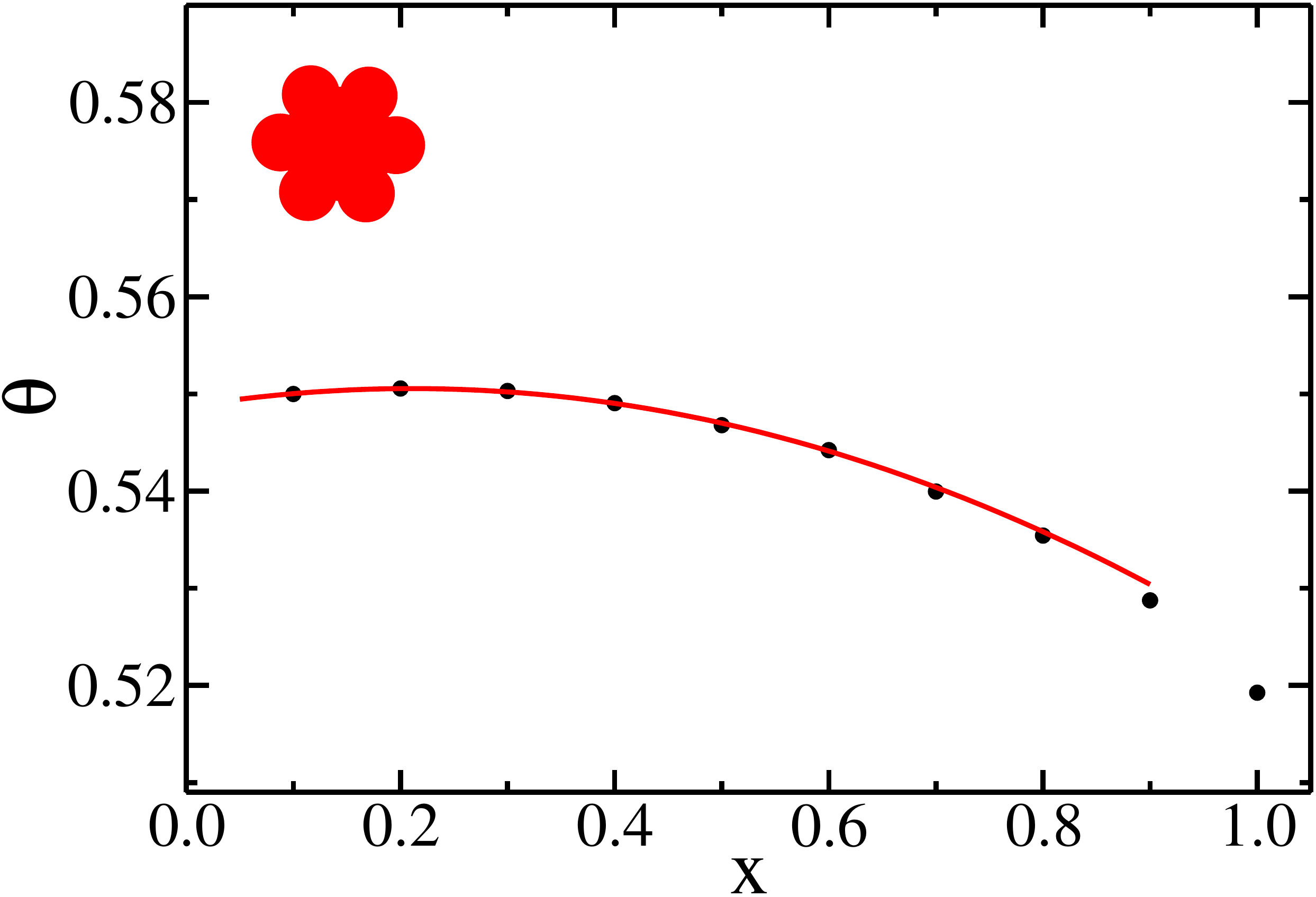}
}
\vspace{0.05\columnwidth}
\centerline{%
\includegraphics[width=0.45\columnwidth]{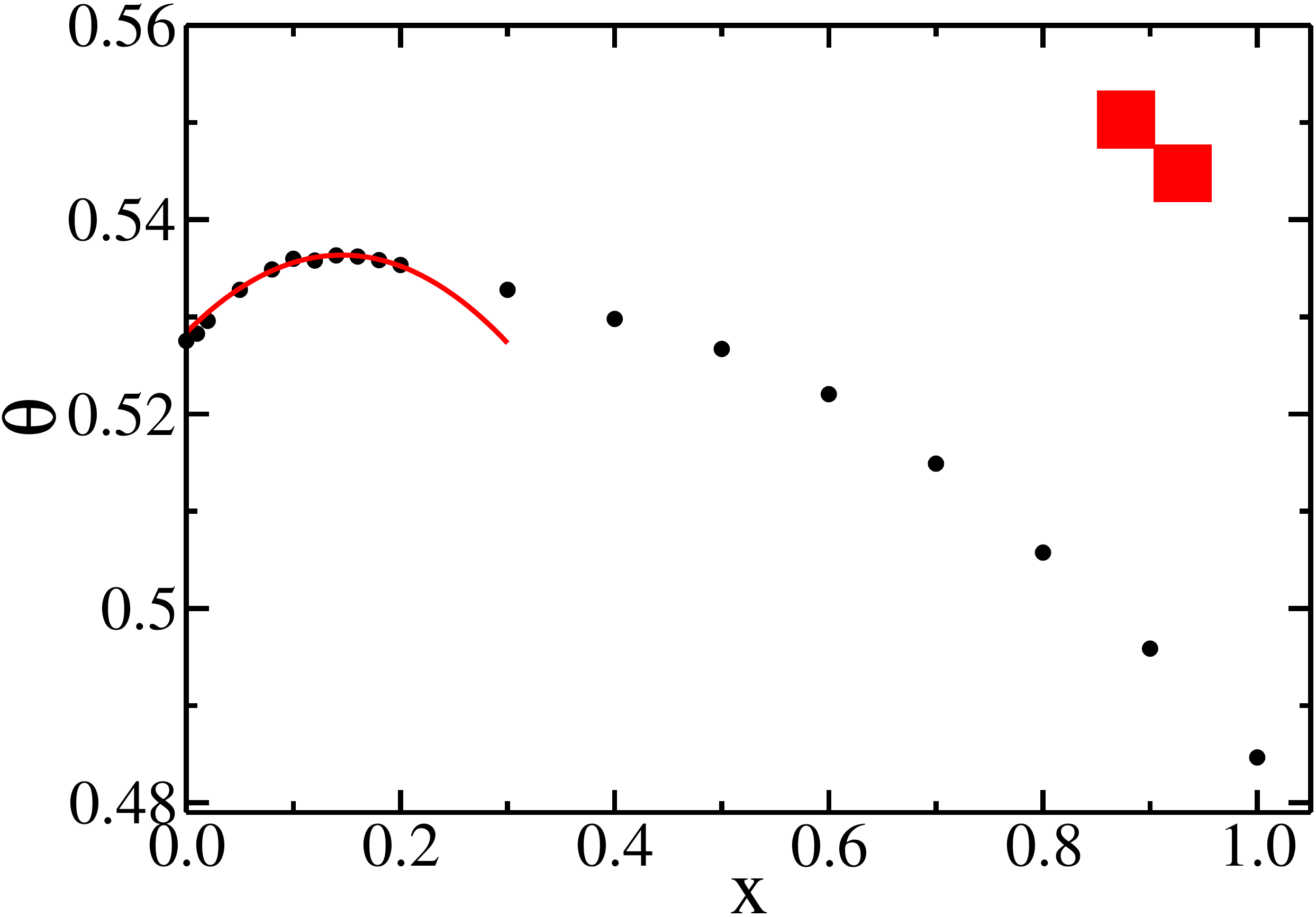}
}
\caption{(Color online) Saturated packing fraction dependence on parameter $x$ for trimers, tetramers, hexamers and square dimers. Dots represents data from numerical simulations. Error bars are smaller than symbol sizes. Solid red lines are quadratic fits near maxima of $\theta_\mathrm{max}$. Data for dimers were taken from \cite{bib:Ciesla2014dim}. The fits are: $\theta_{2} = 0.5483 + 0.12089 x - 0.11751 x^2$, $\theta_{3} = 0.54933 + 0.088299 x - 0.091899 x^2$, $\theta_{4} = 0.54835 + 0.033967 x - 0.050194 x^2$, $\theta_{6} = 0.54867 + 0.017815 x - 0.042357 x^2$, $\theta_{2s} = 0.52839 + 0.10979 x - 0.37802 x^2$.}
\label{fig:q}
\end{figure}

Obtained packing fractions are presented in Fig.~\ref{fig:q}.%
For all shapes the maximal packing fraction is reached for $x \in [0,1]$. This confirms reasoning presented in \cite{bib:Vigil1989} that anisotropy prefers parallel alignment of particles for large $t$, which causes increase of packing fraction, but, on the other hand, at the beginning of RSA the anisotropic particle blocks significantly more space than its surface area, which lowers the packing fraction. Thus, the optimum is reached for a small anisotropy. Fitting a quadratic function to the data near the maximum allows us to estimate an optimal anisotropy, which in our case is determined by parameter $x$, and the value of the highest possible packing fraction. For convenience these data are collected together in the Table~\ref{tab:results}.
\begin{table}[htb]
\caption{\ Maximal possible saturated packing fractions and corresponding values of parameter $x$ for which they are reached. The statistical error of $\theta_\mathrm{max}$ does not exceed $0.0005$.}
\label{tab:results}
\begin{tabular*}{0.5\textwidth}{@{\extracolsep{\fill}}ccc}
\hline
shape & $x$ & $\theta_\mathrm{max}$ \\
\hline
dimer \cite{bib:Ciesla2014dim}       &  0.5098   &  0.5793   \\
trimer       &  0.4804   &  0.5705   \\
tetramer     &  0.3384   &  0.5541   \\
hexamer      &  0.2103   &  0.5505   \\
square dimer &  0.1452   &  0.5364   \\
\hline
\end{tabular*}
\end{table}
\section{Discussion}
\begin{figure}[htb]
\centerline{%
\includegraphics[width=0.7\columnwidth]{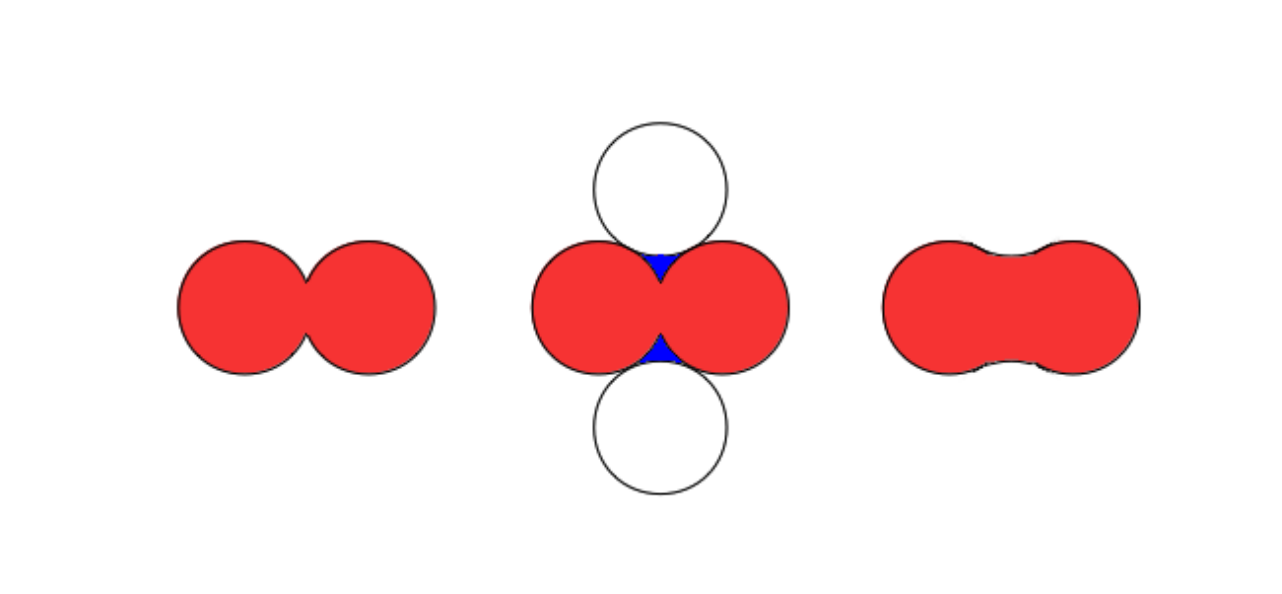}
}
\vspace{0.05\columnwidth}
\centerline{%
\includegraphics[width=0.45\columnwidth]{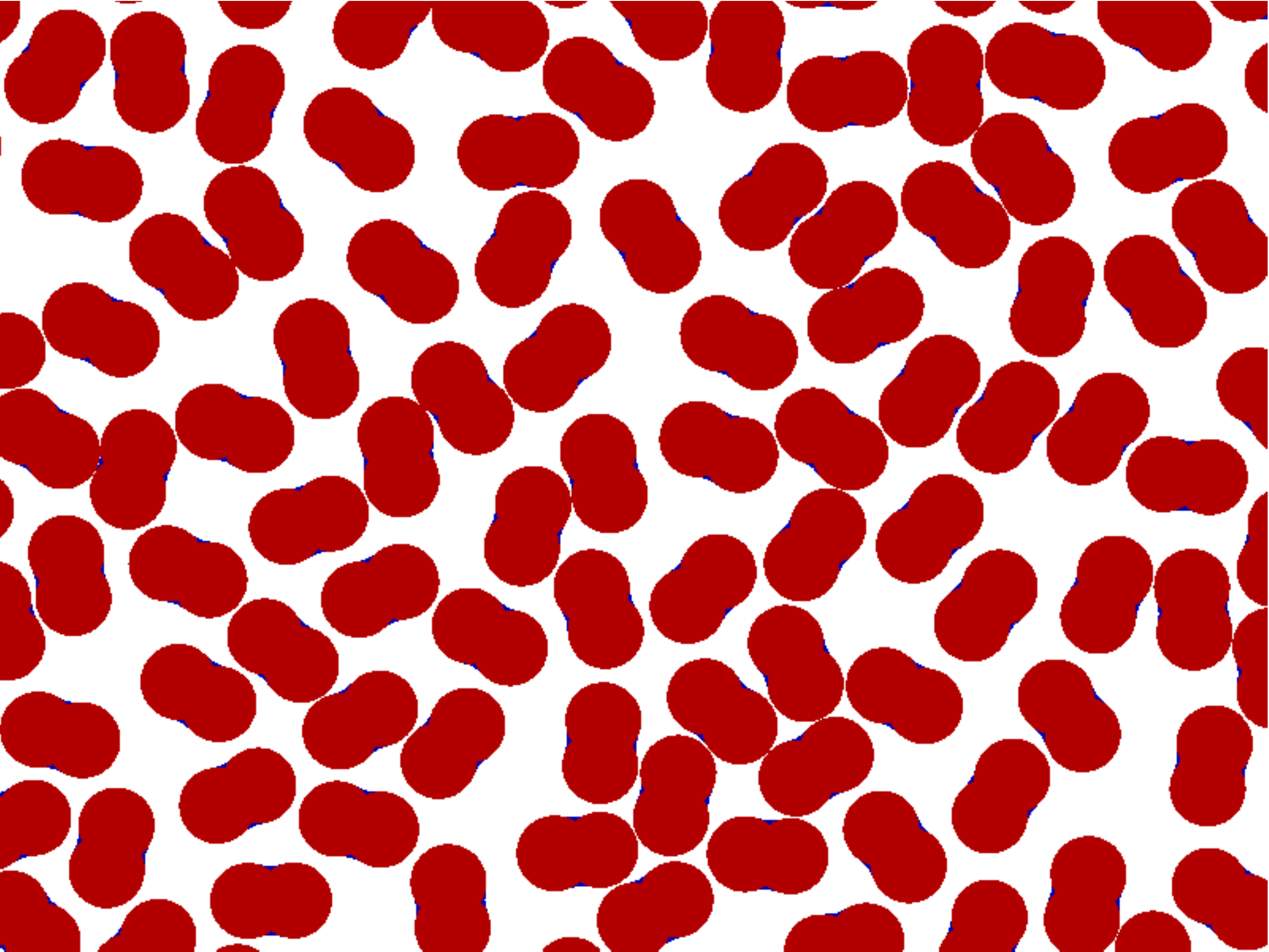}
\hspace{0.05\columnwidth}
\includegraphics[width=0.45\columnwidth]{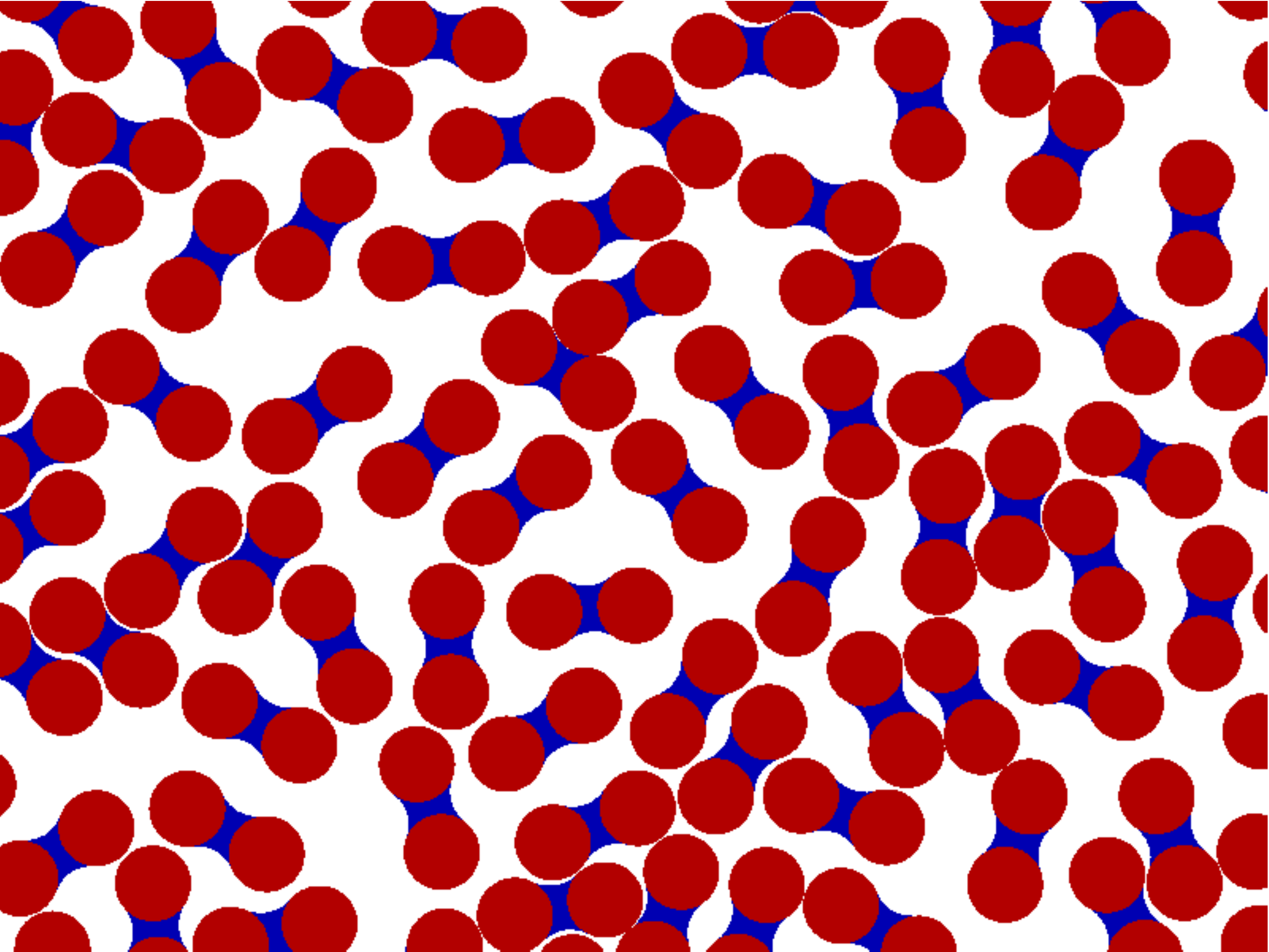}
}
\caption{(Color online) Top panel: the transition from generalized dimer to a shape with larger surface area and the same blocking area. Bottom panels: packings at the end of simulation bulid of such shapes for $x=0.6$ (left) and $x=1.2$ (right).}
\label{fig:bettershapes}
\end{figure}

For all shapes we studied, the maximal possible packing fraction is smaller than for generalized dimers \cite{bib:Ciesla2014dim}. The more complex shape type, the lower the value of the maximum packing fraction. The lowest values are obtain for square dimer. In this last case the $\theta_\mathrm{max}$ is smaller than for rectangles ($0.55$) \cite{bib:Vigil1989, bib:Viot1992} as well as for disks ($0.5470$) \cite{bib:Zhang2013}. In all cases it is smaller than value for ellipsoids or spherocylinders ($0.583$) reported in \cite{bib:Viot1992}. However, these dimers, trimers tetramers and hexamers do not have the best possible shape to reach highest random packing fraction. As  illustrated in Fig.~\ref{fig:bettershapes}
the small blue areas in both sides of a dimer cannot be occupied by any other particle. It means that this surface is blocked but it is not counted in the packing fraction. There is a direct projection between packing composed of dimers and rightmost particles in Fig.~\ref{fig:bettershapes}. Therefore saturated random packing fraction for this shape will be slightly higher than for dimers. Similar reasoning can be performed for trimers, tetramers and hexamers. Note that this reasoning is valid also for $x>1$, however, too large a value of $x$ can destroy the mentioned projection between two types of shapes. For example for dimer the projection is valid for $x \le \sqrt{2}$.
\par
\begin{figure}[htb]
\centerline{%
\includegraphics[width=0.45\columnwidth]{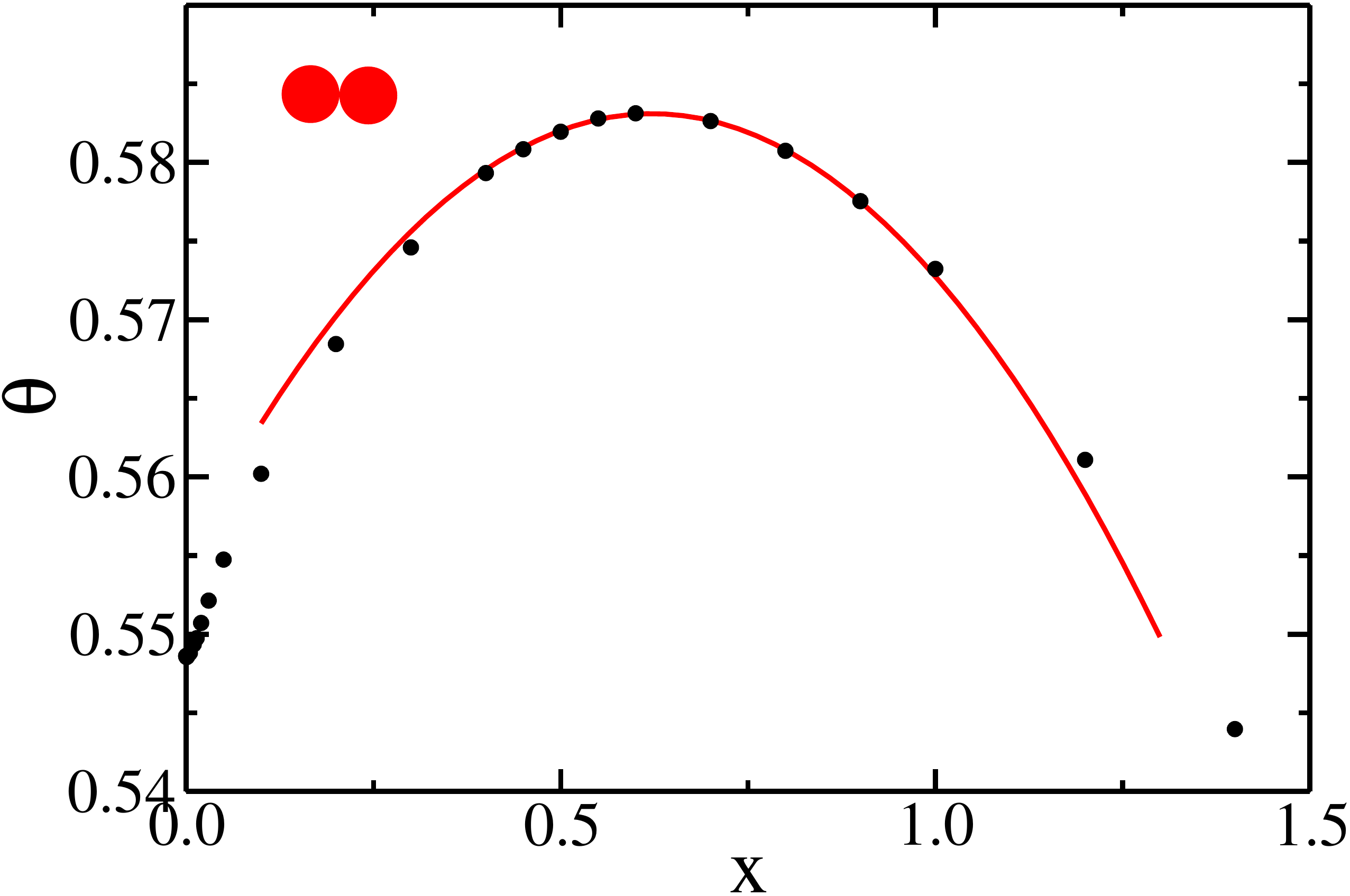}
\hspace{0.05\columnwidth}
\includegraphics[width=0.45\columnwidth]{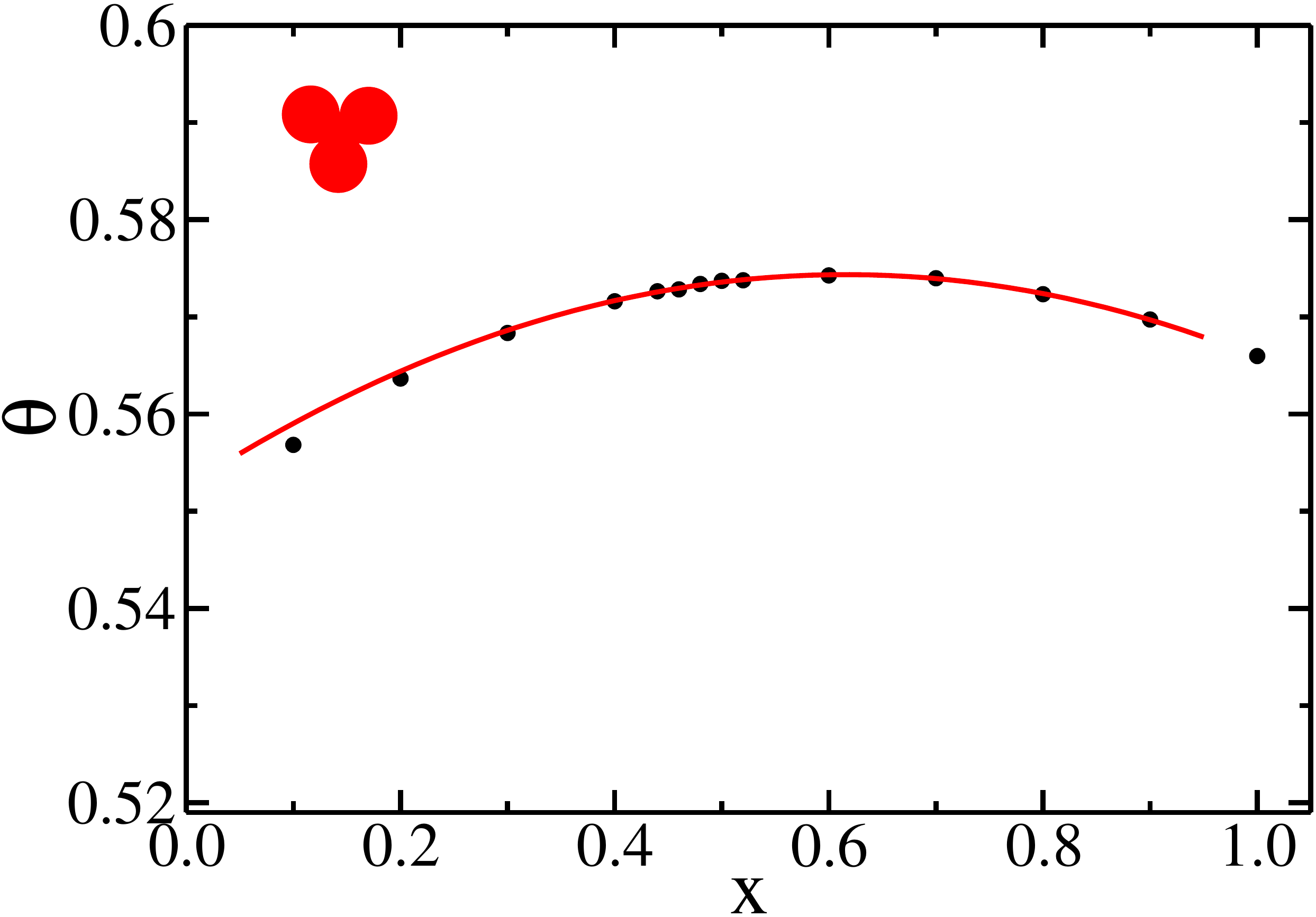}
}
\vspace{0.05\columnwidth}
\centerline{%
\includegraphics[width=0.45\columnwidth]{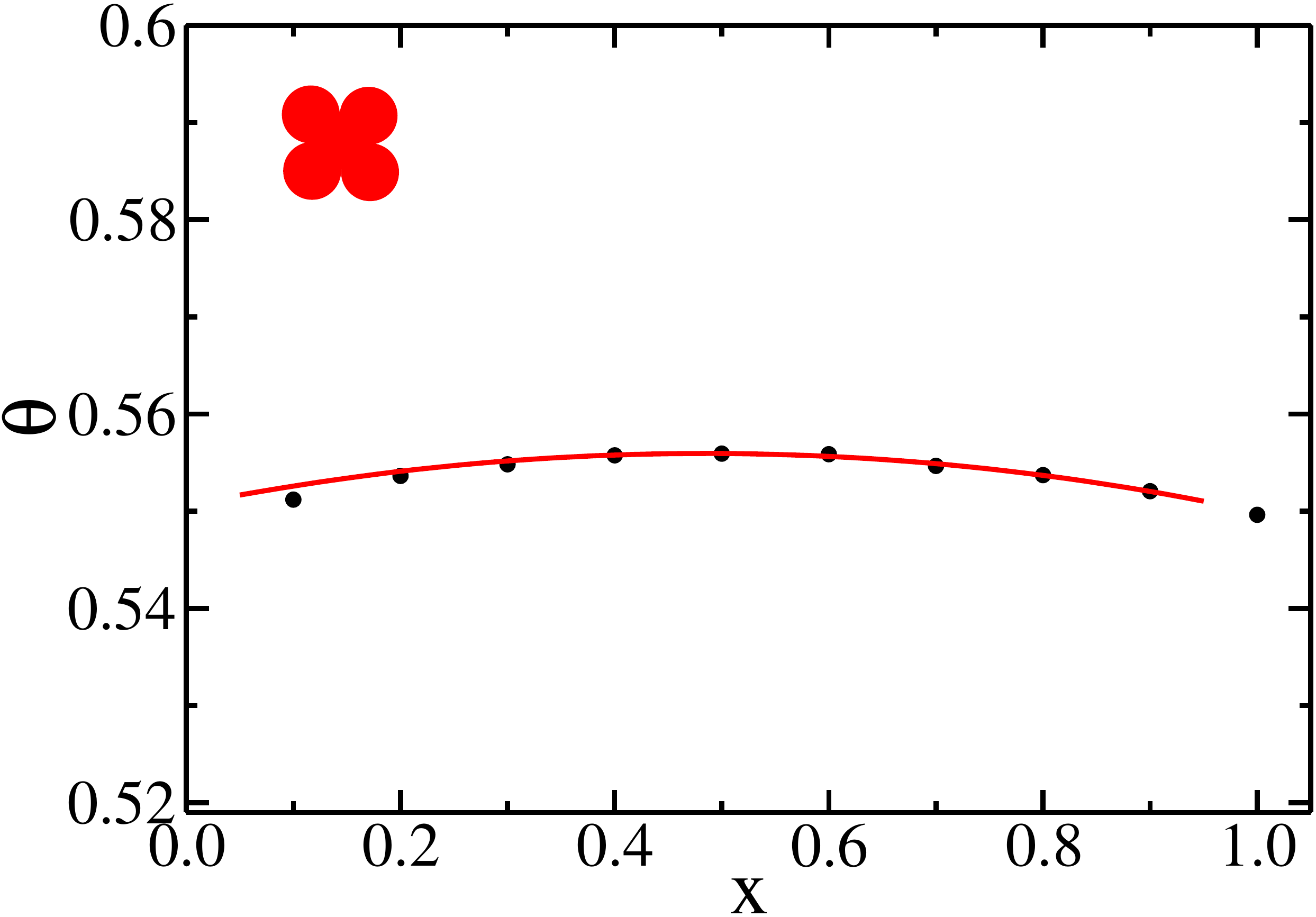}
\hspace{0.05\columnwidth}
\includegraphics[width=0.45\columnwidth]{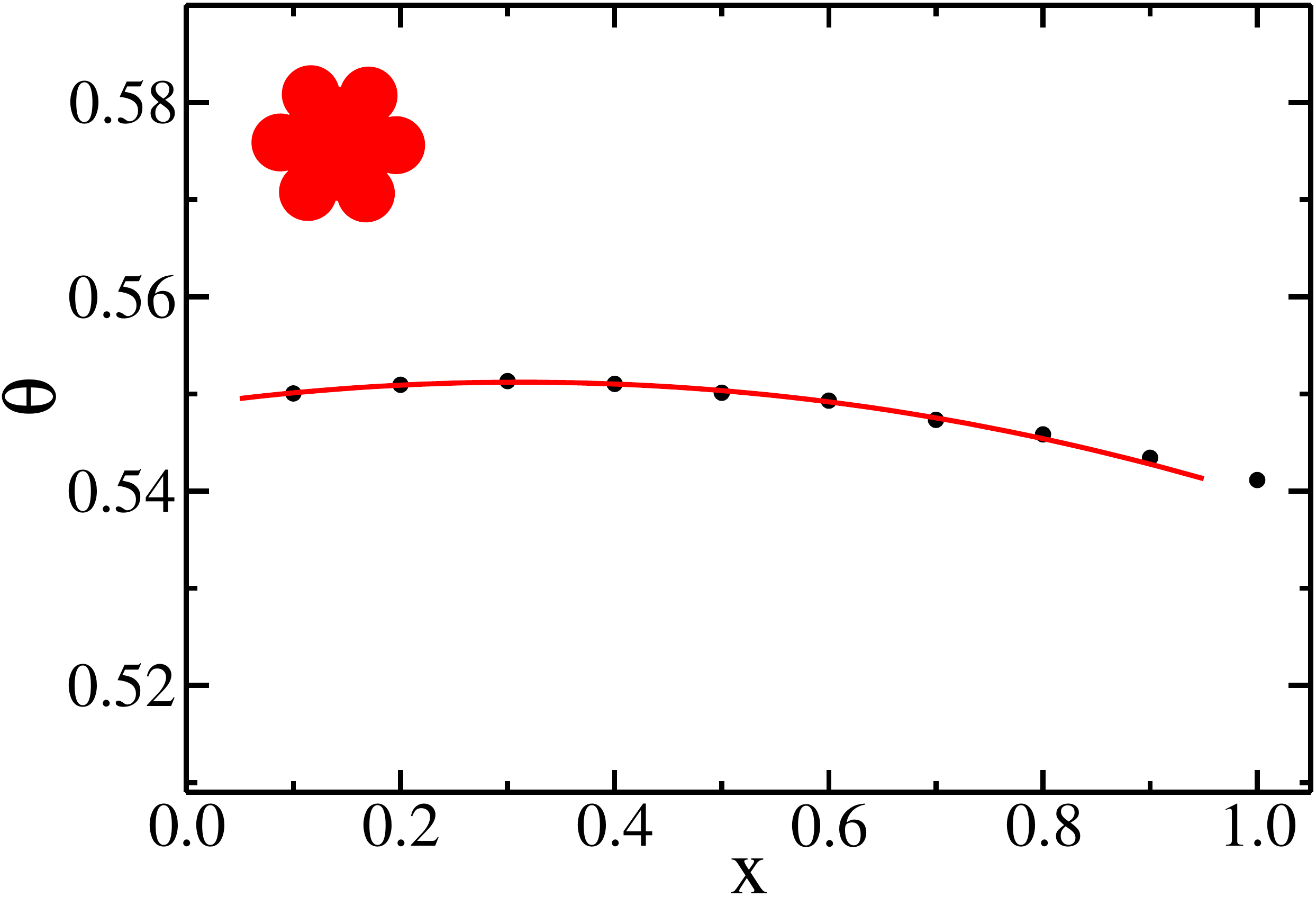}
}
\caption{(Color online) Saturated packing fraction dependence on parameter $x$ for dimers, trimers, tetramers and hexamers with additional blocked regions taken into account (``Smoothed" shapes). Dots represents data from numerical simulations. Error bars are smaller than symbol sizes. Solid red lines are quadratic fits near maxima of $\theta_\mathrm{max}$. Data for dimers were taken from \cite{bib:Ciesla2014dim}. The fits are $\theta_{2} = 0.55513 + 0.089974 x - 0.072349 x^2$, $\theta_{3} = 0.55253 + 0.070875 x - 0.057558 x^2$, $\theta_{4} = 0.55062 + 0.021978 x - 0.022673 x^2$, $\theta_{6} = 0.54883 + 0.015239 x - 0.024414 x^2$.}
\label{fig:qn}
\end{figure}

The surface area of one of these additional regions is
\begin{equation}
S_\mathrm{ad}(x) = \left\{
\begin{array}{c c}
 x \left(\sqrt{4 - x^2} -\sqrt{1 - x^2} \right) - 
 \arcsin x & 0 \leq x < 1 \\
 x \sqrt{4 - x^2} -\frac{\pi}{2} & 1 \leq x \le \sqrt{3}
\end{array}
\right.
\end{equation}
Taking this into account and performing the same analysis (see Fig.~\ref{fig:qn}) as in Sec.~\ref{sec:packing} the following values are obtained (see Table~\ref{tab:conclusions}). 
Note that in case of dimers with $x>1$ the additional surface does not compensate a smaller number of particles in the packing so the packing fraction lowers with growing $x$. Therefore we focused on $x \le 1$ in the case of trimers, tetramers and hexamers.
\begin{table}[htb]
\caption{\ Maximal possible saturated packing fractions and corresponding values of parameter $x$ for which they are reached. Here the additional blocked regions were taken into account (``smoothed'' shapes).  The statistical error of $\theta_\mathrm{max}$ does not exceed $0.0005$.}
\label{tab:conclusions}
\begin{tabular*}{0.5\textwidth}{@{\extracolsep{\fill}}ccc}
\hline
shape & $x$ & $\theta_\mathrm{max}$ \\
\hline
dimer        &  0.6347   &  0.5833   \\
trimer       &  0.6146   &  0.5744   \\
tetramer     &  0.4969   &  0.5560   \\
hexamer      &  0.3036   &  0.5513   \\
\hline
\end{tabular*}
\end{table}
The calculated packing fractions are bigger, but the difference is less than $0.005$ and becomes smaller with growing number of disks in particle. On the other hand, the value obtained for dimeric particle is now similar to the one for ellipsoids and spherocylinders. As expected, the value of $x$ where the packing reaches a maximum is larger when we include the blocked areas in the particles. 
For the case of the smoothed out dimer, we are finding a maximum coverage at $x = 0.6347$, or in other words the aspect ratio $\alpha$ (ratio of maximum dimensions in the two directions) is $1.6347$.  This compares with a maximum for other shapes as given in Table~\ref{tab:summary}.
\begin{table}[htbp]
\caption{\ Comparison of the maximal coverages for various oblong shapes.  $\alpha$ is the aspect ratio, equalling $ 1 + x$ for dimers.  Numbers in parentheses are errors in the last digit. }
\label{tab:summary}
\begin{tabular*}{0.5\textwidth}{@{\extracolsep{\fill}}cccc}
\hline
shape &$\alpha$ & $\theta_\mathrm{max}$ & Ref. \\
\hline
rectangle       &  1.618      &  0.553(1)     & \cite{bib:Viot1992} \\
dimer           &  1.5098     &  0.5793(1)    & \cite{bib:Ciesla2014dim}  \\
ellipse         &  2.0        &  0.583(1)  & \cite{bib:Viot1992} \\
spherocylinder  &  1.75       &  0.583(1)  & \cite{bib:Viot1992} \\
smoothed dimer    &  1.6347     &  0.5833(5) & this work \\
\hline
\end{tabular*}
\end{table}
Evidently, by making the smoothed concave dimer shape, one is able to reach a coverage comparable to the maximum of certain convex shapes (the ellipse and spherocylinder).  This is because in RSA there is a tendency for particles of moderate aspect ratio to adsorb efficiently in both parallel and perpendicular directions.  Having an indentation in the sides of the particles allows one of the disk ends of a neighboring particle to occasionally fit in nicely, and thus increasing the overall coverage.  With longer aspect ratios, large open spaces form which lowers to overall coverage.
\par
\begin{figure}[htb]
\vspace{0.05\columnwidth}
\centerline{%
\includegraphics[width=0.7\columnwidth]{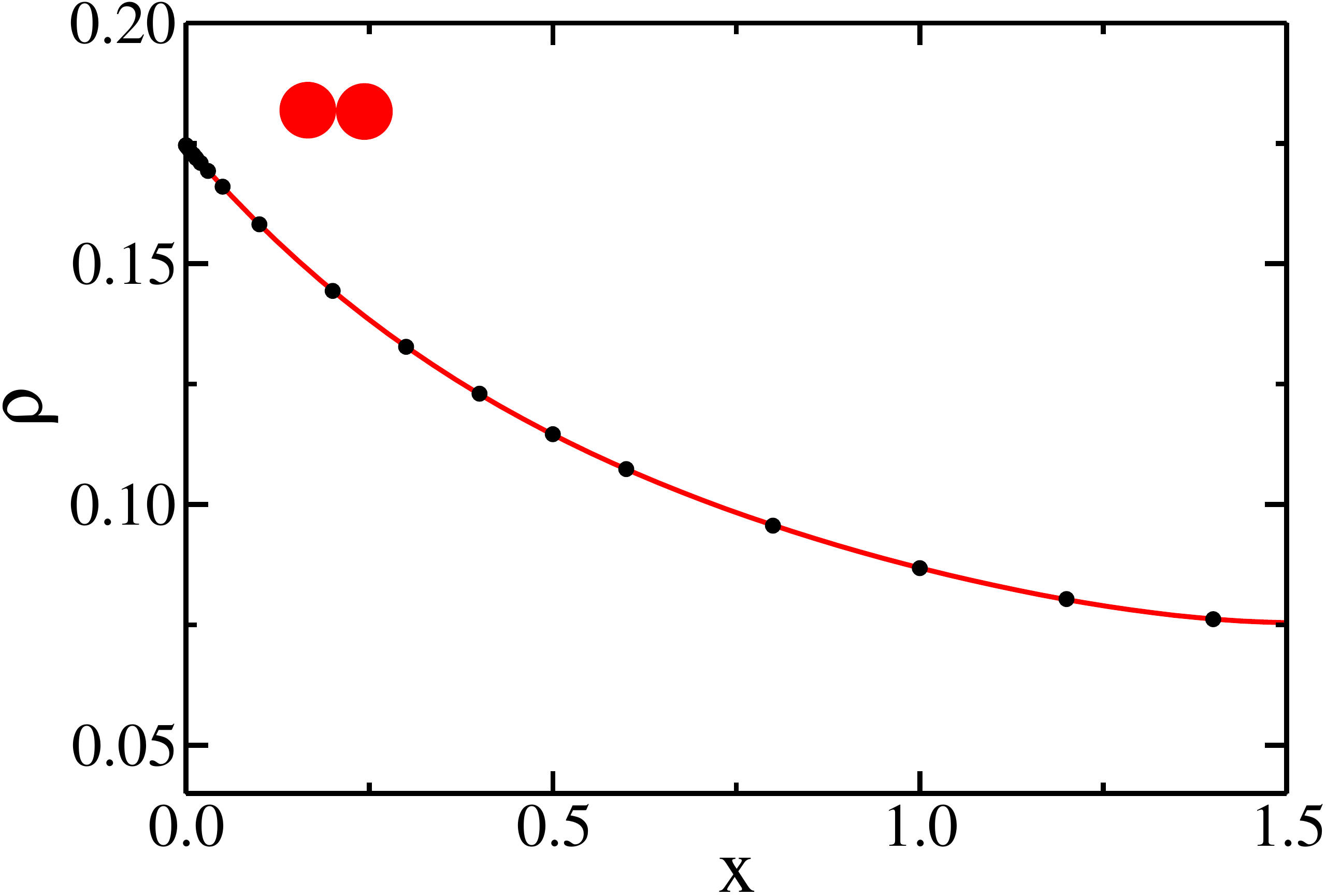}
}
\caption{(Color online) The dependence of the surface density of dimers (both standard and smoothed) on parameter $x$. Dots represents data from numerical simulations. Error bars are smaller than symbol sizes. Solid red lines is a $4^{th}$ order fit of $\rho(x) = y = 0.17452 - 0.17742 x + 0.14968 x^2 - 0.079263 x^3 + 0.019315 x^4$.}
\label{fig:rho}
\end{figure}

The quadratic fits given in captions of Figs.~\ref{fig:q} and \ref{fig:qn} are valid only in a close neighborhood of the maximum coverage. If someone is interested in finding saturated packing fraction of dimers for arbitrary $x$ it is better to use the $\rho(x)$ dependence presented in the caption of  Fig.~\ref{fig:rho}. 
The saturated packing fraction is then given by:
\begin{equation}
\theta_\mathrm{max}(x) = S(x) \rho(x),
\end{equation}
where $S(x) = S_2(x)$ for standard dimer and $S(x) = S_2(x) + 2S_\mathrm{ad}(x)$ for smoothed one. 
\section{Summary}
Several anisotropic and concave shapes of particles were analyzed in terms of maximal possible random packing fraction. It was found that the highest packing fraction is obtained for particle built of two disks with additional blocking area (see Fig.~\ref{fig:bettershapes}), where the distance between their centers is $2 \times 0.6347$ which interestingly is more than $20\%$ larger than in the case without the extended blocking region. The saturated random packing fraction for such a shape is $0.5833 \pm 0.0001$ and is at the level of the highest of previously reported values for other particle shapes.  Future studies of higher precision can show which of the high coverage shapes (smoothed dimer, ellipse, or spherocylinder or other oblong object) is the absolute best.

\section{Acknowledgments}
The authors thank Jim Evans for comments on the paper.
G.~Paj\c{a}k acknowledges support of Cracovian Consortium `$,\!,$Materia-Energia-Przysz\l{}o\'s\'c'' im. Mariana Smoluchowskiego' within the KNOW grant. 

\end{document}